\documentclass[%
aip,
amsmath,amssymb,
reprint,%
]{revtex4-2}

\usepackage{graphicx}
\usepackage{dcolumn}
\usepackage{bm}

\usepackage[utf8]{inputenc}
\usepackage[T1]{fontenc}
\usepackage{mathptmx}
\usepackage{etoolbox}
\usepackage{mhchem}
\usepackage{soul}

\usepackage{blindtext}

\makeatletter
\def\@email#1#2{%
	\endgroup
	\patchcmd{\titleblock@produce}
	{\frontmatter@RRAPformat}
	{\frontmatter@RRAPformat{\produce@RRAP{*#1\href{mailto:#2}{#2}}}\frontmatter@RRAPformat}
	{}{}
}%
\makeatother

\draft
\begin{document}

	
	
	\title{Exploring Cesium-Tellurium phase space via high-throughput calculations beyond semi-local density-functional theory }
	
	
	
	\author{Holger-Dietrich Sa{\ss}nick}
	\affiliation{
		Carl von Ossietzky Universit\"at Oldenburg, Physics Department, D-26129 Oldenburg, Germany
	}
	
	\author{Caterina Cocchi}%
 \email{caterina.cocchi@uni-oldenburg.de}
	\affiliation{
		Carl von Ossietzky Universit\"at Oldenburg, Physics Department, D-26129 Oldenburg, Germany
	}
	\affiliation{
		Humboldt-Universit\"at zu Berlin, Physics Department and IRIS Adlershof, D-12489 Berlin, Germany
	}
	
	
	\date{\today}
	
	\begin{abstract}
		Boosted by the relentless increase of available computational resources, high-throughput calculations based on first principles methods have become a powerful tool to screen a huge range of materials.
		The backbone of these studies are well-structured and reproducible workflows efficiently returning the desired properties given chemical compositions and atomic arrangements as sole input.
		Herein, we present a new workflow designed to compute the stability and the electronic properties of crystalline materials from density-functional theory using the SCAN approximation for the exchange-correlation potential.
		We show the performance of the developed tool exploring the binary Cs-Te phase space which hosts cesium telluride, a semiconducting material widely used as photocathode in particle accelerators. 
		Starting from a pool of structures retrieved from open computational material databases, we analyze formation energies as a function of relative Cs content and for a few selected crystals, we investigate the band structures and density of states unraveling interconnections among structure, stochiometry, stability, and electronic properties.
		Our study contributes to the ongoing research on alkali-based photocahodes and demonstrates that high-throughput calculations based on state-of-the-art first-principles methods can complement experiments in the search for optimal materials for next-generation electron sources.
		
	\end{abstract}
	
	
	\maketitle 
	
	\section{Introduction}
	
	The search for advanced materials for electron sources has recently become a hot topic well beyond the traditional fields of accelerator physics and electron microscopy.~\cite{dowe+10nimpra,musu+10rsi,musu+18nimpra} 
	Ultrabright electron beams produced by high-yield photocathodes are necessary, among others, for the operation of novel instrumentation such as free electron lasers, ultrafast scattering, and radiation detectors.~\cite{bouc+09nimpra,chen+20prab,mcco21book}
	In view of optimizing the performance of these applications, it is crucial to identify novel materials with optimal characteristics and sustainable growth and operational conditions.
	For these tasks, traditional trial-and-error procedures are expensive and ineffective.
	New approaches from computational material science, exploiting the results of \textit{ab initio} simulation to gain insight into the material properties~\cite{cocc-sass21micromachines,anto+21am,loui+21natm,marz+21natm} have recently emerged as viable alternatives and/or complements to this empirical approach. 
	The availability of open-access quantum material databases since the last few years~\cite{curt+12cms,jain+13aplm,saal+13jom,drax-sche19jpm,tali+20sd} represents a valuable resource to find potential candidates for target applications.
	Likewise, the development of automated workflows for density-functional theory (DFT) calculations~\cite{goss+18cms,pizzi+18mrs,yaku+21cms} in connection with machine learning techniques~\cite{schl+19jpm,chib-coud20aplm,chen+20cms} has opened even broader perspectives to identify and design materials with suitable characteristics as photocathodes.~\cite{anto+21am} 
	
	Despite their great potential, available schemes for high-throughput screening inevitably suffer from the intrinsic limitations of DFT with semi-local approximations for the exchange-correlation (xc) potential.
	The choice of these functionals is traditionally driven by the need to optimize the computational effort when exploring large configurational spaces including several thousands of systems. 
	Underestimation of band-gaps is the most serious consequence of this approach but  additional drawbacks manifest themselves in the formation energies as well as in the description of the density of states,\cite{sahn+20jpcl,ran+21jpcl} especially when defects are present in the systems.~\cite{dahl+21ees}
	While the consolidated awareness of the problems of semi-local DFT enables the application of effective workarounds,~\cite{anto+21am} such corrections are typically empirical and do not cover the most serious cases in which DFT fails reproducing even qualitatively the electronic structure of the materials. 
	On the other hand, many-body perturbation theory ($GW$ approximation and Bethe-Salpeter equation),~\cite{onid+02rmp} the state-of-the-art method to compute electronic and optical excitations in solids, cannot be conveniently applied to high-throughput calculations yet.
	The reason for this is not only in the much higher computational costs but especially in the non-standard, system-dependent convergence process of the numerous computational parameters involved that finally determine the reliability of the final results.
	
	The intense efforts dedicated in the last decades to the development of hybrid xc functionals for DFT~\cite{b3lyp,adam-baro99jcp,hse03} have ultimately led to approximations that deliver band gaps in excellent agreement with experimental references for a wide range of materials.~\cite{garz-scus16jpcl,borl+19jctc}
	An alternative approach that has been explored more recently is based on meta-GGA, the rung directly above the generalized gradient approximation (GGA) in the ``Jacob’s ladder" of xc functionals.~\cite{burk12jcp}
	Corresponding implementations, including the SCAN functional,~\cite{sun+15prl} are computationally less expensive than hybrid functionals but generally lead to superior accuracy compared to the GGA.\cite{jana+18jcp,borl+19jctc}
	In particular, we recently showed that SCAN is able to predict band-gaps of Cs-based antimonides and tellurides in very good agreement with many-body perturbation theory results.~\cite{sass-cocc21es}
	Our understanding of this behavior is based on the availability in those materials of $s$- and $p$-like bands close to the frontier.
	Benchmark studies on elemental solids such as diamond, silicon, and germanium confirm this trend.~\cite{pisc+20cms,yao-kana17jcp}
	
	In this work, we present an automated workflow for DFT calculations employing the SCAN functional to explore the stability and the electronic structure of cesium-telluride materials. 
	These systems, although largely used as electron sources in many particle accelerators around the globe,~\cite{kong+95jap,dowe+10nimpra} are still poorly characterized and understood from a fundamental perspective. 
	This lack of knowledge, in turn, inhibits the possibility to enhance the performance of the resulting photocathodes in a controlled way.
	The proposed approach, built upon the AiiDA infrastructure~\cite{pizzi+18mrs} and supporting calculations performed with the CP2K package,~\cite{kueh+20jcp} is interfaced with an efficient data-mining routine that identifies suitable structures from computational material databases and inputs them for the subsequent DFT calculations. 
	The developed workflow includes steps for structural optimization and self-consistent calculations to determine	the energetics and the electronic structure of stable stoichiometries, which can coexist with the nominally grown \ce{Cs2Te} phase. The subsequent analysis of band-gaps and projected densities of states is aimed at establishing correlations with the crystal structure and the chemical composition of the materials. 
	The obtained results can be used as an advanced input to three-step model for photoemission model.~\cite{berg-spic64pr} 
	Moreover, the presented computational scheme is ready to be interfaced with algorithms for crystal structure prediction as well as with machine learning approaches.
	
	This paper is organized as follows: In Sec.~\ref{sec:methods} we introduce the developed high-throughput workflow (Sec.~\ref{ssec:flow}) and summarize the details of DFT calculations (Sec.~\ref{ssec:cp2k}). 
	In Sec.~\ref{sec:results}, we apply the workflow to explore the cesium-tellurium phase space, introducing the adopted approach to select the initial data pool (Sec.~\ref{ssec:pool}) and subsequently investigating structural stability (Sec.~\ref{ssec:hull}) and its connection with the electronic properties of the materials (Sec.~\ref{ssec:elect}), with specific focus on the experimental stoichiometry (Sec.~\ref{ssec:2:1}).
	Finally, we summarize our results, we present our conclusions, and propose an outlook on prospect follow-up work (Sec.~\ref{sec:conclusions}).

	\section{Computational Methods}\label{sec:methods}
	
	To perform high-throughput calculations exploring a large configurational space in an efficient and reliable way, an automatized workflow is needed to setup and manage the computational procedure as well as to handle and process the output.
	Our development stands upon the existing open-source python library AiiDA,~\cite{hube+20sd, uhri+21cms} which provides a robust and yet flexible computational infrastructure for high-throughput DFT calculations.
	AiiDA has been specifically designed to automatize complex computational workflows, to manage large amounts of data, and to promote sharing scientific results ensuring data provenance.
	To this end, all input and output files as well as all calculation steps are stored as nodes in a database, thereby enabling easy back-tracing and reproduction of results.
	While the core functions are included in the main library, interfaces to specific software packages are handled via plugins.
	In our implementation, we use the official AiiDA plugin for CP2K~\cite{aiida_cp2k21github} as a base, and supply it with additional, purposely implemented features.
	These developments are part of a custom library that will be released as a standalone, open-source package in the near future.
	In the following, we outline the implemented workflow, describing its parts and specifying how they interact with each other. 
	A general description of the AiiDA infrastructure of  workflows can be found elsewhere.~\cite{pizzi+16cms, hube+20sd, uhri+21cms}
	
	\subsection{The high-throughput workflow}\label{ssec:flow}
	
	\begin{figure}
		\includegraphics[scale=1.0]{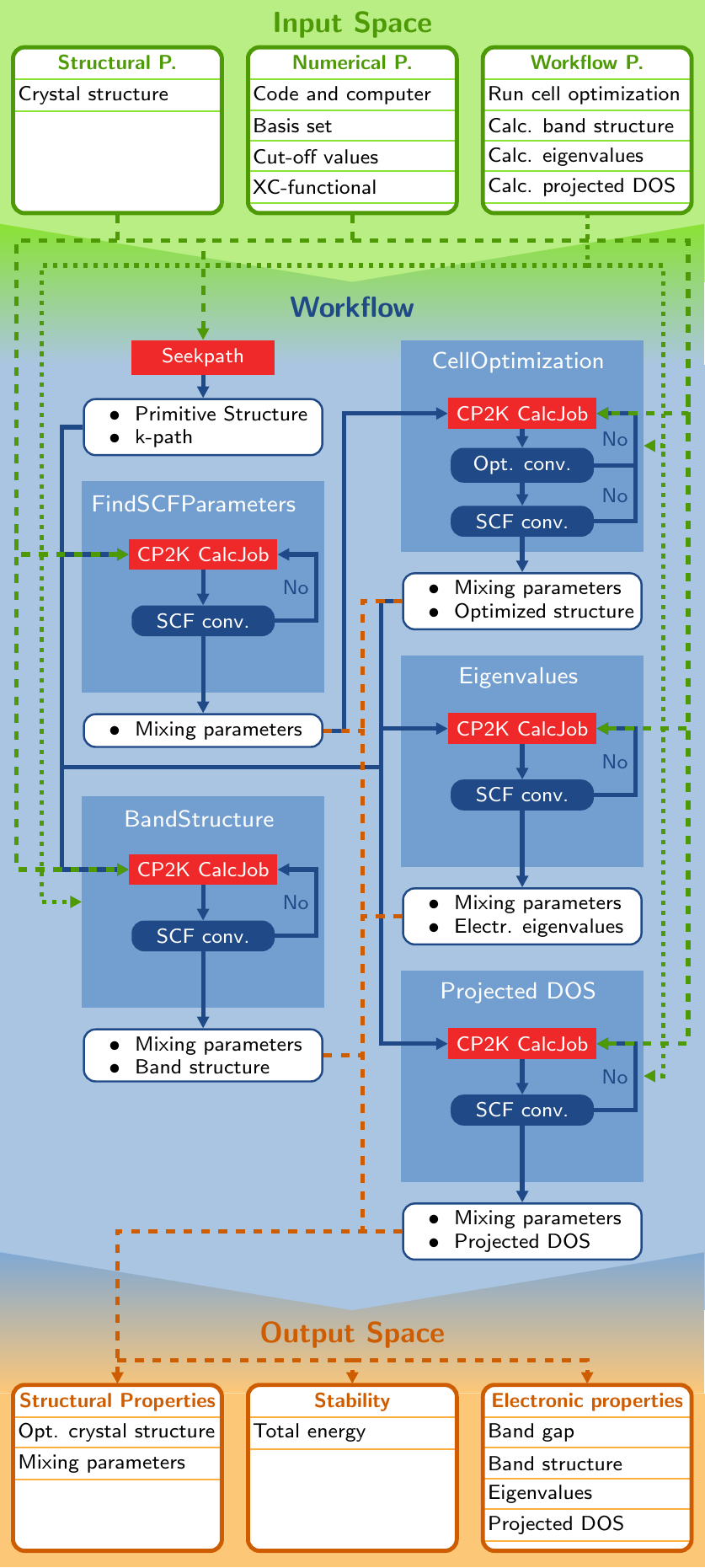}%
		\caption{\label{fig:workflow} Schematic overview of the developed computational workflow. The green area on top represents the space all relevant input parameters; the blue area in the middle includes the steps of the workflow, where all relevant sub-processes are indicated within rectangular fields; the orange area shows calculated properties that are stored as output nodes in the database.
			Green and orange dashed arrows show the data transfer from the input space into the workflow and from the workflow to the output space, respectively.
			Dark blue arrows denote transfer of data between the sub-processes defining dependencies among them.
			Green dotted arrows show the transfer of control parameters.}
	\end{figure}
	
	The developed workflow, schematically represented in Fig.~\ref{fig:workflow}, is implemented as an AiiDA \texttt{WorkChain}~\cite{hube+20sd} object hierarchically calling sub-processes (blue and red rectangular boxes with corners) to perform different tasks.
	There are three different types of input parameters in the input space shown in the green area on the top of Fig.~\ref{fig:workflow}: \textit{Structural parameters} define the crystal structure and its properties, \textit{numerical parameters} specify the calculation accuracy, and \textit{workflow parameters} manage the functionality of the workflow.
	The input space contains only the most relevant parameters, thus keeping the usability as simple as possible.
	All remaining settings are given by a protocol-file that can be generated and adapted for special use cases.
	
	After the definition of all input parameters, the actual workflow begins, consisting of a logical sequence of predefined stages.
	The individual workflow stages are depicted as rectangular filled boxes in Fig.~\ref{fig:workflow}, whereby AiiDA objects such as \texttt{CalcJobs} or \texttt{CalcFunctions} and \texttt{WorkChains} (see Ref.~\citenum{hube+20sd} for details) are shown in red and light blue, respectively.
	After each DFT calculation, process handlers, represented in Fig.~\ref{fig:workflow} as filled dark blue boxes with rounded corners, are set in place to check for specific problems, e.g., whether the cell optimization or the convergence of self-consistent field (SCF) cycles was successful.
	Dark blue arrows indicate exchanged data between different stages of the workflow, thus highlighting their interdependencies.
	
	The first stage of the workflow, preliminary to any DFT calculations, consists of the determination of the space group of the considered crystal structure, its primitive unit cell, as well as a standardized \textbf{k}-path throughout the first Brillouin zone.
	To do so, the two external libraries \texttt{SeeK-path}~\cite{hinu+17cms} and \texttt{spglib}~\cite{togo-tana18arxiv} are used.
	In the first \texttt{WorkChain} for DFT calculations, the mixing and smearing parameters for convergence of the Kohn-Sham equations~\cite{kohn-sham65pr} are determined.
	A SCF calculation is performed iterating through different mixing parameters and density mixing schemes, including Broyden- and Pulay-mixing~\cite{baar-vand11jcp} as well as Kerker damping as implemented in CP2K.~\cite{kueh+20jcp}
	For each parameter set, four different electronic smearing temperatures are tried using a Fermi-Dirac distribution.
	During all following stages, only the mixing-parameter is adjusted while the mixing-scheme and smearing temperature are kept fixed.
	For the case of super cell calculations of materials with a band-gap the more efficient orbital transformation method is also implemented.~\cite{vand-hutt03jcp}
	
	Next, the unit cell parameters and the atomic positions are optimized minimizing interatomic forces and internal pressure down to a specified threshold, using the previously set mixing parameters to converge the Kohn-Sham equations.
	Similar to the previous step, the unit cell optimization is implemented as an iterative procedure.
	To ensure structural optimization within an extended phase space hosting numerous systems with different potential energy surfaces, the \texttt{WorkChain} iterates through a set of diversified optimization algorithms and parameters.
	Adopting initially the Broyden-Fletcher-Goldfarb-Shanno algorithm and starting from a very large trust radius combined with a small number of steps, the trust radius is consecutively reduced while the number of optimization iterations is increased.
	As a last resort, the much slower but more stable conjugated gradient method is used.
	After each structural optimization run, the convergence of the crystal geometry and of the last SCF-cycle are checked.
	
	The final stages of the workflow, currently implemented in three different \texttt{WorkChains}, run in parallel and calculate the electronic properties using the optimized structure and the final mixing parameters.
	Kohn-Sham energy eigenvalues are calculated using a denser mesh with \textbf{k}-points separated by 0.075~\AA$^{-1}$ from each other.
	The calculation of the band structure is performed along a predefined \textbf{k}-path derived via the \texttt{SeeK-path} library.
	The projected density of states is obtained simulating the system in a supercell, such that the cell length in each dimension is larger than 25~\AA, and sampling its Brillouin zone at the $\Gamma$-point only.
We adopt this approach as implemented in CP2K in order to enable an efficient screening of the key features of the DOS in the considered materials. Comparison with reference calculations from Ref.~\citenum{sass-cocc21es} reported in the Supplementary Material confirms the validity of this choice. 
	
	The output space (orange area at the bottom of Fig.~\ref{fig:workflow}) of the workflow is variably set according to the previous steps, with the final mixing parameters and the total energy being returned and stored in any case.
	The total energy of the crystal is always extracted from the calculation using the tightest parameters.
	Finally, all relevant output errors are identified and back-traced through the hierarchy of the workflow.
	
	\subsection{Details of CP2K Calculations}\label{ssec:cp2k}
	
	All DFT calculations performed within the workflow desctribed above are run using the CP2K package, implementing Gaussian and plane wave basis-set schemes.~\cite{vand+05cpc}
	In the analysis of the Cs-Te phase space, we have chosen triple-$\zeta$ valence basis sets including two polarization functions (MOLOPT-TZV2P) for both elements, a plane wave cut-off of 550~Ry, as well as relative cutoff value of 100~Ry to ensure sufficient numerical accuracy. 
	To describe core electrons appropriately, we have used dual-space pseudopotentials (GTH-pseudopotentials)~\cite{goed+96prb} represented by Gaussian functions and optimized for the SCAN functional.~\cite{hutt21github}

	\section{Results}\label{sec:results}
	
	We apply the high-throughput workflow introduced above to study the configurational space of binary Cs-Te crystals.
	Cesium telluride is a photocathode material~\cite{powe+73prb} typically grown on a substrate via physical vapour deposition.~\cite{gaow+19prab}
	The obtained crystal structures and phases are generally not known with precision but it is assumed that several different phases with different stoichiometry can form and coexist in polycrystalline samples.~\cite{powe+73prb}
	As optimal photoemission properties are associated with smooth and homogeneous materials, the knowledge of crystal structure and stoichiometry of the phases formed during growth is essential.
	Furthermore, for the application of this material as an electron source, access to the electronic properties is of fundamental relevance too.
	Unfortunately, this body of information can be hardly accessed experimentally, especially in light of the known issues with material growth and its degradation.~\cite{gaow+19prab}
	DFT can greatly help in this regard, providing unrivaled insight into the structure-property relations of photocathode materials, as demonstrated for \ce{Cs2Te} itself~\cite{sass-cocc21es,terd+12prb} as well for (multi)alkali antimonides.~\cite{kala+10jpcs,alay-shau11jms,murt+16bms,cocc+18jpcm,cocc+19sr,cocc20pssrrl,amad+21jpcm,cocc-sass21micromachines}
	
	\subsection{Initial dataset}\label{ssec:pool}
	
	\begin{figure*}
		\includegraphics[scale=1.0]{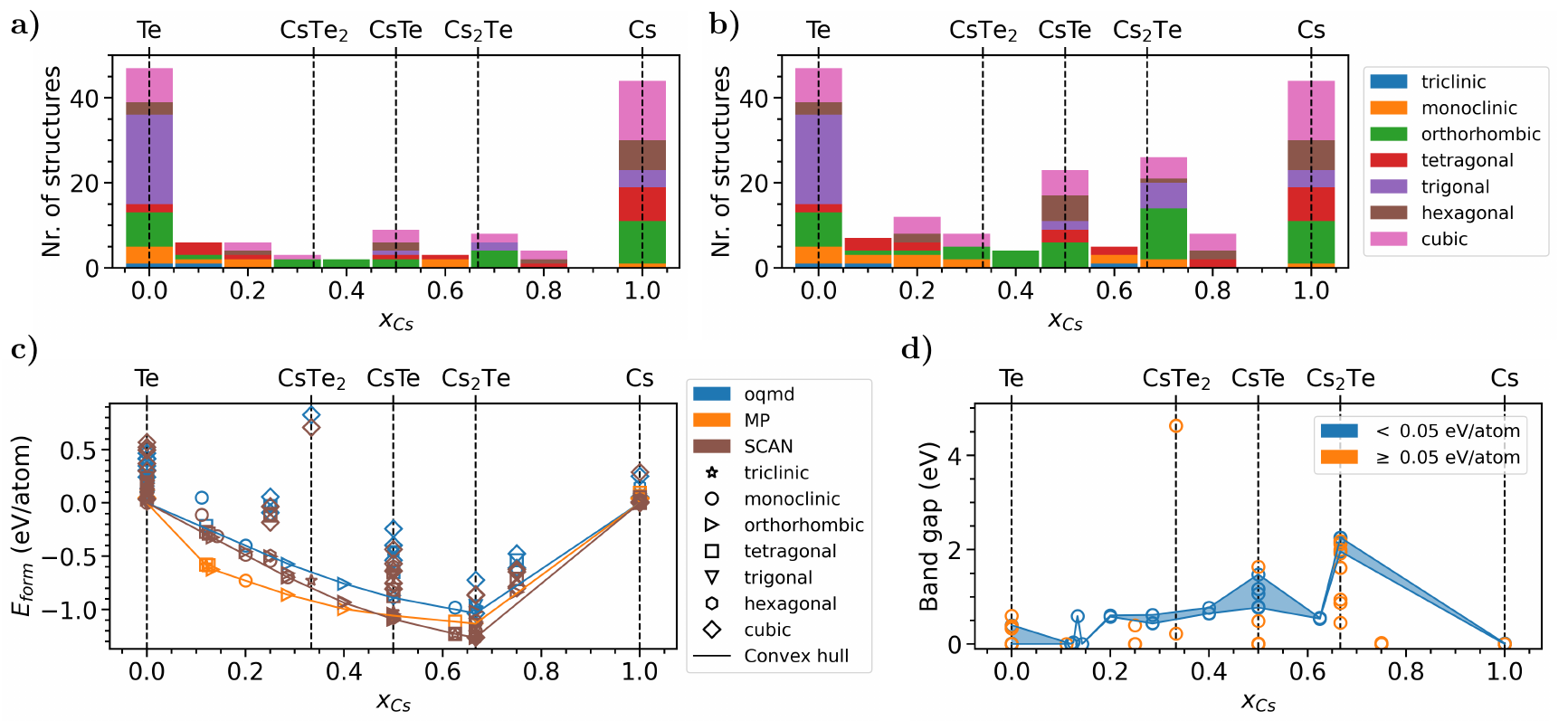}%
		\caption{\label{fig:Cs-Te_overview} Top panel: Overview of the Cs-Te structures mined from MP and OQMD a) without and b) with additional structures generated from chemically analogous materials upon exchange of anions and cations; the experimental structure with composition 5:3 is included as well in the dataset. Bottom panel: c) Formation energies ($E_{form}$) including the convex hull and d) band gaps of stable structures \textit{versus} relative Cs concentration.}%
	\end{figure*}
	
	As a starting point for our analysis, we consider all crystal structures exclusively containing Cs and Te elements stored in the Materials Project (MP) database~\cite{jain+13aplm} and in the Open Quantum Materials Database (OQMD).~\cite{saal+13jom}
Elemental phases of Cs and Te available in the aforementioned databases are included in the initial dataset as a point of reference for subsequent calculations of formation energies and for the analysis of the phase diagram of the binary systems.
	In addition, the crystal structure of monoclinic Cs$_\mathrm{5}$Te$_\mathrm{3}$~\cite{sche-boet91} is manually added to the dataset, as it is missing in either database.
	To minimize the overlap between the entries from MP and OQMD, we have applied the \textit{F-Fingerprint} method~\cite{ogan-vall09jcp} to filter out duplicate structures.
	This procedure results in the 132 structures visualized in Fig.~\ref{fig:Cs-Te_overview}a).
	
	A careful inspection of Fig.~\ref{fig:Cs-Te_overview}a) reveals the unbalance of the obtained dataset towards the elemental phases.
	In order to reduce the weight of the latter, in the pool we include additional binary crystals derived from materials formed by cations and anions belonging to the same chemical groups as Cs and Te, respectively.
	Specifically, we consider crystal structures containing K and Rb as cations and Se and Po as anions in place of Cs and Te, respectively.
	The unit cell dimensions are scaled according to the ratio between the ionic radii of the substituted elements.
	The inclusion of this additional set of 52 structures significantly impacts on the phase space incrementing the relative amount of binary compounds in comparison with the elemental phases (see Fig.~\ref{fig:Cs-Te_overview} b). 
	The new configurational space includes 184 crystal structures in total.
	
	All generated crystal structures are taken as input by the high-throughput workflow described in Sec.~\ref{sec:methods}.
	DFT calculations on the 132 crystal structures mined from MP and OQMD converged without problems, while among the 52 additionally generated structures 4 had to be excluded for numerical issues.
	All crystal structures that successfully passed this step in the workflow are compared with each other once again using the \textit{F-fingerprint} method filtering out a total of 19 duplicated structures and consequently resulting in 161 successfully calculated independent crystal structures.

	\subsection{Stability and convex hull}\label{ssec:hull}
	
	The first step in the analysis of the Cs-Te phase space consists of the assessment of structural stability.
	This quantity is evaluated from the formation energy defined as
	\begin{eqnarray}
		E_{form}(\mathrm{Cs}_x\mathrm{Te}_{1-x}) = E(\mathrm{Cs}_x\mathrm{Te}_{1-x})\nonumber\\
		- [xE(\mathrm{Cs}) + (1-x)E(\mathrm{Te}) ],
		\label{eq:form_e}
	\end{eqnarray}
	where $E(\mathrm{Cs}_x\mathrm{Te}_{1-x})$ is the energy per atom of the binary compound while $E(\mathrm{Cs})$ and $E(\mathrm{Te})$ are its counterparts for the most stable elemental phases.
	These values are obtained from DFT total energies, and, as such, they do not include zero-point energy contributions nor thermal effects.
	Positive values of the formation energy are indicative of unstable structures.
	
	The formation energies of all calculated phases using SCAN are plotted in Fig.~\ref{fig:Cs-Te_overview}c), where the corresponding values extracted from MP and OQMD are displayed as well.
	The results stored in both databases are obtained using the DFT package VASP~\cite{kres-furt96prb} employing the PBE parametrization~\cite{pbe} of the GGA xc functional.~\cite{jain+13aplm, saal+13jom}
	The discrepancy between MP and OQMD datasets, in spite of the analogous computational parameters adopted to generate them, originates from an empirical correction for Te-containing compounds applied to the formation energies in MP.~\cite{wang+21sr}
	The convex hull shown in Fig.~\ref{fig:Cs-Te_overview}c) is obtained for each dataset (OQMD, MP, and our own SCAN results) by connecting the formation energies of the most stable phase for each composition including linear combinations of phases with different compositions.
	Comparing our results obtained with the SCAN functional with the convex hull resulting from the OQMD dataset, we notice systematically lower formation energies for the former, although qualitatively the trends of formation energy \textit{versus} relative Cs content remain the same.
	This finding, consistent with similar analysis performed on binary Cs-Sb crystal in previous work,~\cite{cocc-sass21micromachines} can be ascribed to the improved functional adopted in our calculations.
	Considering now our results against the MP dataset, we notice a qualitative difference.
	The formation energies stored in MP are systematically lower than our data points towards tellurium-rich phases (see Fig.~\ref{fig:Cs-Te_overview}c) as the magnitude of the aforementioned correction scheme depends on the tellurium content of the phases.~\cite{wang+21sr}
	Nonetheless, the minimum of the convex hull in all three datasets is found for a composition with 2:1 Cs:Te relative content. 
	The absolute minimum ($E_{form}^{min}=-$1.26~eV/atom) is obtained with SCAN while with PBE we find $E_{form}^{min}=-$1.13~eV/atom for the MP dataset and $E_{form}^{min}=-$1.04~eV/atom for the OQMD one.
	
	It is worth recalling that the phase diagram of Cs-Te materials has already been constructed based on experimental data.~\cite{pham+15calphad}
	Although a quantitative correlation between these results and DFT data is hindered by the many variables in play (experimental conditions, presence of defects or impurities, etc.), still, such a comparison can provide an insightful point of reference in our analysis.
	The synthesized phases with the space group / chemical formula $P2_1/c$~[14] / CsTe$_\mathrm{4}$,~\cite{boet-kret85} $Cmcm$~[63] / Cs$_\mathrm{2}$Te$_\mathrm{5}$,~\cite{boet+kret82} $Cmc2_1$~[36] / Cs$_\mathrm{2}$Te$_\mathrm{3}$,~\cite{boet80jlcm} $Pbam$ [55] / CsTe,~\cite{debo-cord95jac} $C2/m$ [12] / Cs$_\mathrm{5}$Te$_\mathrm{3}$,~\cite{sche-boet91} and $Pnma$~[62] / Cs$_\mathrm{2}$Te~\cite{sche-boet91} are located at the convex hull in our calculations.
	For the Cs$_\mathrm{5}$Te$_\mathrm{3}$ and Cs$_\mathrm{2}$Te, enthalpies of formation $-(942.2 \pm 8.3)$~kJ$\cdot$mol$^{-1}$ and $-(362.9 \pm 2.9)$~kJ$\cdot$mol$^{-1}$, respectively, have been derived from solution calorimetry experiments~\cite{debo-cord97} at ambient conditions.
	Transformed into eV/atom, we obtain the experimental reference values of $-(1.22 \pm 0.01)$~eV/atom and  $-(1.25 \pm 0.01)$~eV/atom for Cs$_\mathrm{5}$Te$_\mathrm{3}$ and Cs$_\mathrm{2}$Te, respectively.	
	Looking at the formation energies of the most stable phases for these two materials for the different datasets, we obtain $-0.98$~eV/atom and $-1.04$~eV/atom for OQMD, $-1.11$~eV/atom and $-1.13$~eV/atom for MP and $-1.23$~eV/atom and $-1.26$~eV/atom for our calculations, respectively.
	In both cases, the formation energy calculated with the SCAN functional matches the experimental value within its error margin while the other methods give overestimates of these values, leading to underbound structures.
	The very good agreement between our SCAN results and available experimental data highlights the overall improvement provided by our approach in the prediction of structural stability for this material class compared to GGA and empirical correction schemes.
	It is worth mentioning that additional contributions coming from temperature and pressure are not yet present in our calculations.
	Their inclusion can further improve the agreement with experiments and, more importantly, our knowledge of these materials.

	\subsection{Correlation between structural properties and band gaps}\label{ssec:elect}
	
	Turning now to the electronic properties of the investigated set of materials, we analyze the band gap values against the relative Cs content. 
	To further emphasise correlations with the structural stability of the materials, we filter out those phases that exhibit formation energies that are at least 0.05~eV/atom higher than the convex hull and that, as such, are less likely to form under experimental conditions (orange dots in Fig.~\ref{fig:Cs-Te_overview}~d).
	For the remaining phases, above the the convex hull by up to 0.05~eV/atom (blue dots in Fig.~\ref{fig:Cs-Te_overview}~d), a general tendency towards larger band gaps can be noticed upon increasing cesium content. 
	The maximum is reached at the Cs:Te composition with 2:1 ratio, to which the stoichiometry of the known compound \ce{Cs2Te} belongs.~\cite{sche-boet91}
	With the aid of the blue area highlighted in Fig.~\ref{fig:Cs-Te_overview}~d), we can identify the band-gap range with increasing Cs relative content. 
	Although not monotonic, a tendency towards increasing band-gap can be noticed upon larger amounts of Cs, dropping to zero at the elemental phase of the Cs metal.
	Interestingly, the largest variability of band-gap values is noticed at the 1:1 composition, where an energy range from 0.77~eV to 1.48~eV is spanned.
	
	\begin{figure*}
		\includegraphics[scale=1.0]{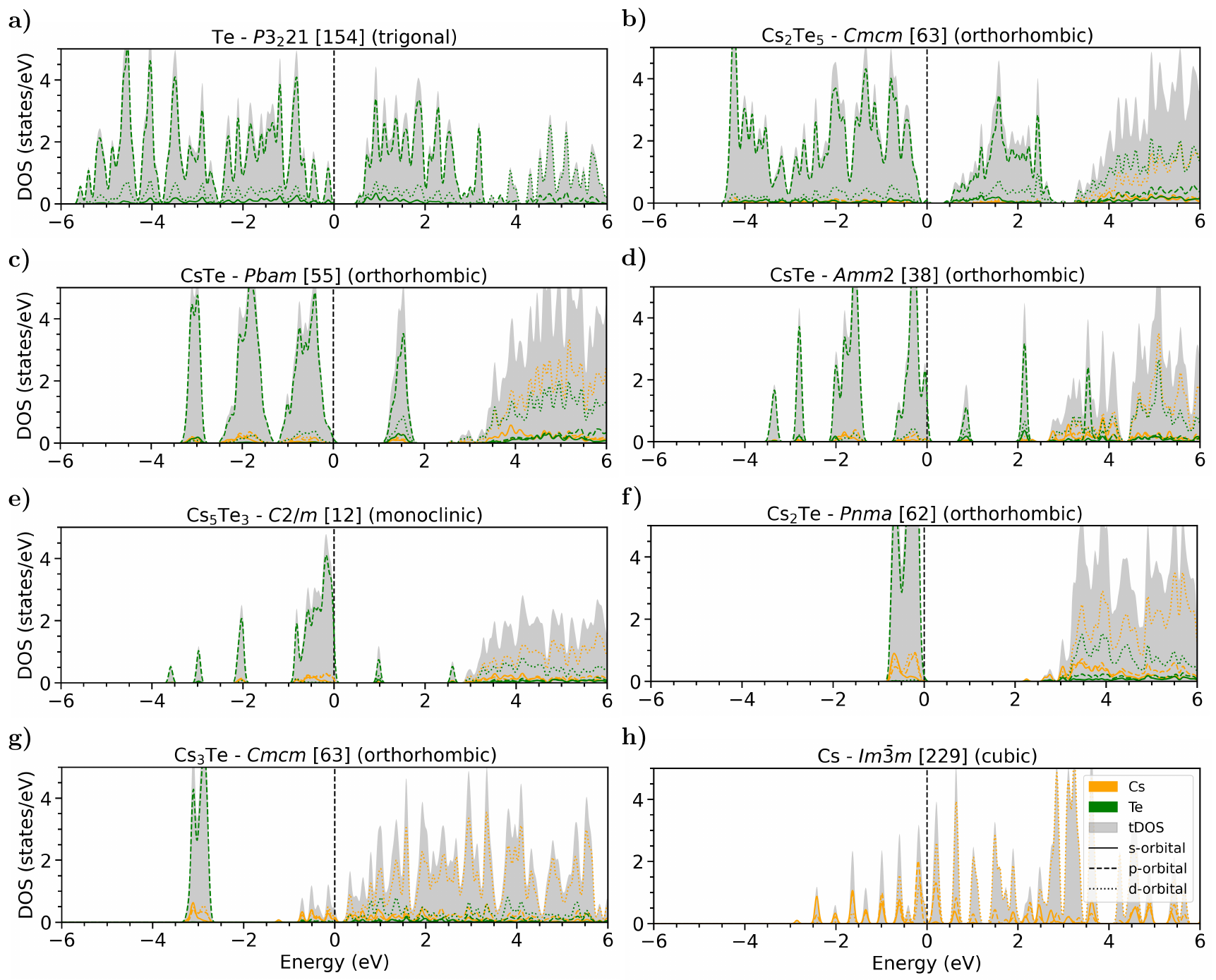}%
		\caption{\label{fig:pdos}Projected density of states (pDOS) of six selected crystal structures with increasing relative Cs-content from a) to f).
			Solid, dashed, dotted, and dash-dotted lines represent contributions from $s$-, $p$-, $d$-, and $f$-orbitals, respectively.
			The valence band maximum is set to 0~eV in each plot.}%
	\end{figure*}
	
	The correlation between chemical composition and electronic properties can be studied in more detail by inspecting the projected density of states (pDOS) of selected phases along the phase diagram, see Fig.~\ref{fig:pdos}.
	In this analysis, we mainly focus on experimentally grown binary materials,~\cite{pham+15calphad} in addition to the elemental solids for comparison.
	The transition from the elemental tellurium phase, characterized by purely covalent bonds, towards the mixed phases exhibiting increasingly ionic bonds, leads to metallic cesium-rich phases up to the elemental phase of Cs (see Fig.~\ref{fig:pdos}), following the trends seen in Fig.~\ref{fig:Cs-Te_overview}~d).
	
	In the elemental Te phase, $p$-orbitals dominate both, the valence region as well as the lowest conduction states, forming two broad bands across the band gap of the material that are clearly visible in the pDOS (Fig.~\ref{fig:pdos}~a).
	Similar features are still present in Te-rich phases, such as in \ce{Cs2Te5} (see Fig.~\ref{fig:pdos}~b), where, however, the energy range of both $p$-orbital dominated valence and conduction bands decreases by a few eV in comparison with elemental tellurium.
	This trend can be directly associated with the presence of chemical bonds of increasing ionic character, which in turn lead to a more pronounced wave-function localization towards Te atoms rather than along the bonds.
	Such a behavior is more pronounced when the relative amount of Cs and Te content becomes equal (Fig.~\ref{fig:pdos}~c-d).
	While valence and conduction states around the gap are still dominated by Te $p$-states, hybridization with Cs $s$- and $p$-orbitals becomes increasingly relevant and hybrid Cs-Te $d$-bands dominate the higher conduction region.
	As the relative Cs content exceeds the Te one (Fig.~\ref{fig:pdos}~e-f), the localized character of the Te $p$-states in the gap region is mostly enhanced.
	This characteristic is directly related to the band-gap increase seen for these compositions in Fig.~\ref{fig:Cs-Te_overview}~d).
	The maximum is achieved at the relative Cs:Te content 2:1, where the number of unoccupied $p$-orbitals of the anion equals the number of occupied $s$-orbitals of the cation.
	Thereby, only the valence bands are dominated by the $p$-orbitals of tellurium while the lowest conduction bands are formed by $s$-orbitals of cesium atoms, see Fig.~\ref{fig:pdos}~f).
	Once surpassing this stoichiometry, the band-gap size drops and the material becomes metallic, see Fig.~\ref{fig:pdos}~g).
	The predominance of Cs content in the considered \ce{Cs3Te} is clear in particular in the lowest conduction bands, which are dominated by the $s$- and $d$-states of the cation.
	Hybridized Te $p$-orbitals are pushed down in energy in the valence region, while in the conduction mainly $d$-states of the anion participates in the bonds with Cs.
	
	Looking back at the formation energies in Fig.~\ref{fig:Cs-Te_overview}c), we notice negative values for Cs:Te ratios larger than 2:1 as an indication of (meta)stable systems. 
	However, the energetic separation from the convex hull is in any case larger than 0.1~eV/atom, implying that the excess of Cs leading to partial occupations at the Fermi energy is detrimental for the energetic stability of the material.
	This line of reasoning is supported by the lack of stable phases beyond the 2:1 Cs:Te ratio in the experimental phase diagrams.~\cite{sang-pelt93jpheq, debo-cord95jac, pham+15calphad}
	
	\subsection{The 2:1 composition of Cs-Te materials}\label{ssec:2:1}
	
Among the variety of structures and compositions analyzed so far, the one with 2:1 relative Cs:Te content covers a prominent role in experimental photocathode research~\cite{dibo+96jap, yuso+17prab, pier+21apl} although different stoichoimetries can coexist in the samples.~\cite{gaow+19prab}
Interestingly, this system is not only at the minimum of the computed convex hull (see Fig.~\ref{fig:Cs-Te_overview}c) but, from the analysis of the pDOS, it also exhibits the largest band gap (Fig.~\ref{fig:pdos}).
	Given the importance of this composition and its characteristics, we deepen our analysis examining all stable phases in our data pool with 2:1 Cs:Te ratio. 
	We rate as stable only those systems with formation energies exceeding the convex hull up to 0.05~eV/atom.
	
	\begin{table}[h!]
		\caption{\label{tab:Cs2Te_phases} Composition (Comp.), space group, distance to the convex hull $\Delta E_{hull}$ and band gap ($E_{gap}$) of Cs-Te materials with Cs:Te composition ratio of 2:1. Only systems with formation energies exceeding the convex hull up to 0.05~eV/atom are considered. For each system, the database source is given: ''m.'' indicates additional structures generated from chemically similar ones mined from the corresponding database.}
		\begin{ruledtabular}
			\begin{tabular}{llddl}
				Comp. & \multicolumn{1}{c}{Space group} & \multicolumn{1}{c}{\mbox{$\Delta E_{hull}$ (eV/at.)}} & \multicolumn{1}{c}{\mbox{$E_{gap}$ (eV)}} & \multicolumn{1}{c}{Source} \\
				\hline
				Cs$_8$Te$_4$    & $P1$ [1]           & 0.000 & 2.26 & OQMD \\
				Cs$_8$Te$_4$    & $Pm$ [6]           & 0.000 & 2.25 & OQMD \\
				Cs$_8$Te$_4$    & $P2_1/c$ [14]      & 0.000 & 2.26 & MP m. \\
				Cs$_8$Te$_4$    & $Pnma$ [62]        & 0.000 & 2.25 & MP \\
				\hline
				Cs$_2$Te    & $R\bar{3}m$ [166]  & 0.002 & 2.14 & OQMD \\
				Cs$_2$Te    & $Fm\bar{3}m$ [225] & 0.002 & 2.14 & OQMD \\
				\hline
				Cs$_{18}$Te$_9$ & $R3$ [146]         & 0.004 & 1.96 & OQMD m. \\
				\hline
				Cs$_4$Te$_2$    & $Cmc2_1$ [36]        & 0.018 & 2.16 & MP m. \\
				Cs$_4$Te$_2$    & $Fdd2$ [43]        & 0.018 & 2.18 & MP m. \\
				Cs$_4$Te$_2$    & $P6_3/mmc$ [194]   & 0.018 & 2.22 & MP m. \\
			\end{tabular}
		\end{ruledtabular}
	\end{table}
	
	The summary of such systems considered in this work is reported in Table~\ref{tab:Cs2Te_phases}.
	To simplify the forthcoming discussion, we consider a representative number of Cs-Te crystals with 2:1 ratio in light of the structural similarities which are also reflected in the electronic characteristics, as elaborated in Sec.~\ref{ssec:elect}. 
	According to the stoichiometry 	of the materials, four groups of structures can be identified.
	The most stable ones, all with formation energies coinciding with the convex hull, are those with composition \ce{Cs8Te4}.
	This group includes the experimentally determined phase with the space group $Pnma$ [62].~\cite{sche-boet91}
	Regardless of their polymorphism, they exhibit very similar band-gap values ranging from 2.25 to 2.26~eV, which are the largest found for this compositional ratio.
	Having a composition of \ce{Cs2Te} we identify a trigonal and cubic phase with formation energies exceeding the convex hull by 2~meV/atom and with a band gap of 2.14~eV in both cases.
	Energetically slightly above these systems, we find the trigonal \ce{Cs18Te9} lattice which is characterized by the smallest band gap (1.96~eV) among the values reported in Table~\ref{tab:Cs2Te_phases}.
	Finally, for the stoichiometry \ce{Cs4Te2} we find the least stable structures with formation energies that are 18~meV/atom above the convex hull.
	For this composition, we find the largest variability of the band gap size with respect to the space group: thee smallest one (2.16~eV) is associated to the orthorhombic lattice (space group $Cmc2_1$ [36]), while the largest one (2.22~eV) is associated with the hexagonal phase belonging to the space group $P6_3/mmc$ [194].
	
	\begin{figure}
		\includegraphics[width=0.48\textwidth]{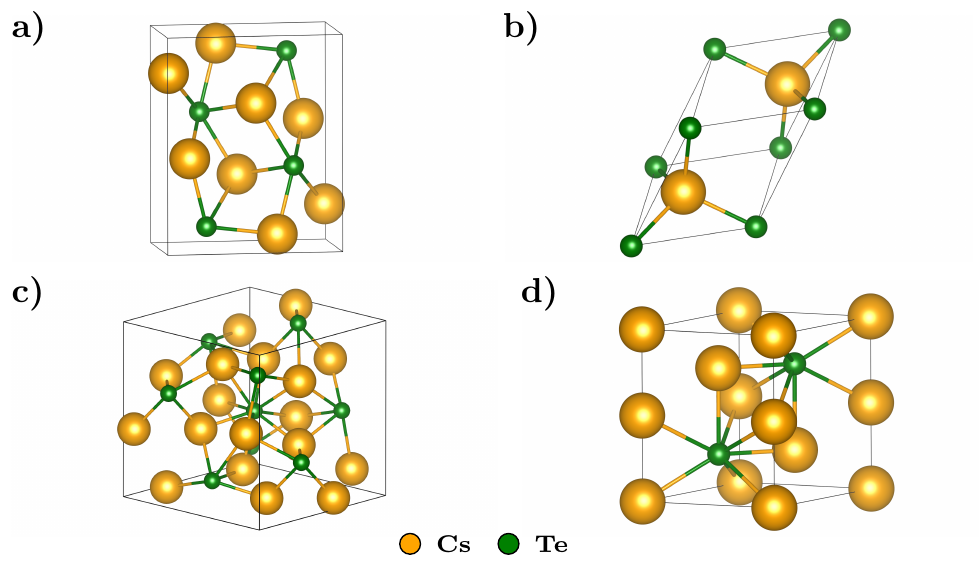}%
		\caption{\label{fig:structures} Ball-and-stick representation of the primitive cells of a) Cs$_8$Te$_4$ with space group $Pnma$ [62], b) Cs$_2$Te$_1$ with space group $Fm\bar{3}m$ [225], c) Cs$_{18}$Te$_9$ with space group $R3$ [146], and d) Cs$_4$Te$_2$ with space group $P6_3/mmc$ [194]. Graphics created with VESTA.~\cite{VESTA}
		}
	\end{figure}
	
	To deepen the analysis on the structure-property relationships of Cs-Te materials with 2:1 compositional ratio, we analyze the band structures of representative materials for each group reported in Table~\ref{tab:Cs2Te_phases}, namely the $Pnma$ [62] orthorhombic phase for \ce{Cs8Te4}, the $Fm\bar{3}m$ [225] cubic phase for \ce{Cs2Te}, and the $P6_3/mmc$ [194] hexagonal phase for \ce{Cs4Te2}.
	As for \ce{Cs18Te9}, we consider the only identified structure belonging to the space group $R3$ [146].
	The corresponding primitive cells are shown in Fig.~\ref{fig:structures}.
	
	\begin{figure*}
		\includegraphics[scale=1.0]{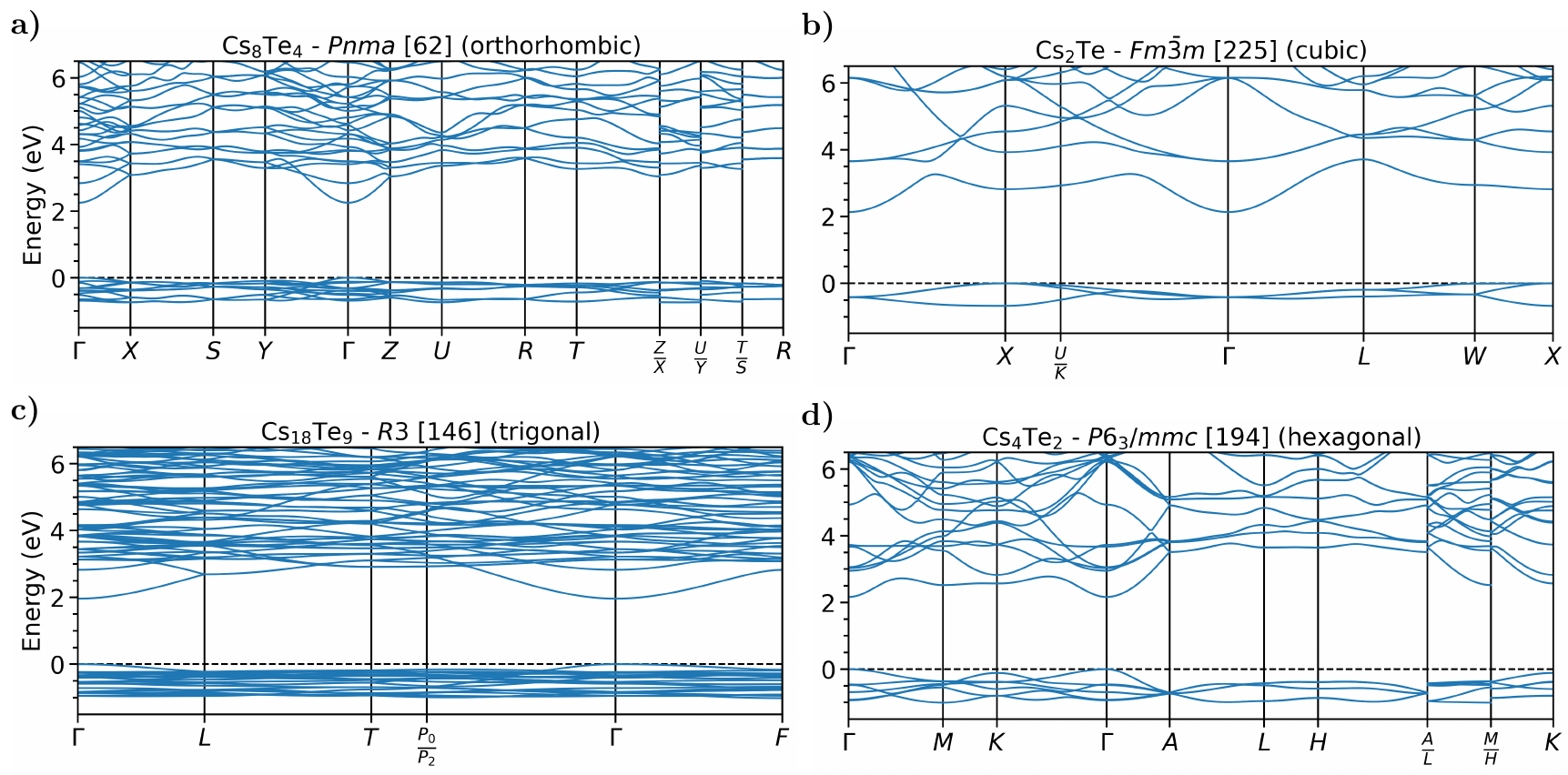}%
		\caption{\label{fig:bands} Band structures of a) Cs$_8$Te$_4$, b) Cs$_2$Te$_1$, c) Cs$_{18}$Te$_9$, and d) Cs$_4$Te$_2$.
			The valence band maximum (VBM) is set to 0~eV in each plot.}%
	\end{figure*}
	
	A bird-eye glance at Fig.~\ref{fig:bands} reveals clear similarities among the systems, irrespective of their stoichiometries and crystal structures.
	Valence bands are rather flat, as expected due to their predominant $p$-character (see Fig.~\ref{fig:pdos} and Ref.~\citenum{sass-cocc21es}), and they extend for a relative narrow energy interval of less than 1~eV.
	The conduction band minimum is located at the $\Gamma$-point in all considered systems and it corresponds to the minimum of a parabolic-like band, consistent with the $s$-character of the corresponding states.
	
	Focusing now on the details, we notice that the band-structure of the cubic phase sticks out with respect to the others (see Fig.~\ref{fig:bands}b).
	Notably, the unit cell of this material (Fig.~\ref{fig:structures}b) exhibits a strong resemblance to the cubic phase of \ce{Cs3Sb}~\cite{sass-cocc21es, cocc-sass21micromachines} with the main difference being the missing atom at site ($\frac{1}{2}, \frac{1}{2}, \frac{1}{2}$), due to the different coordination between Cs and Te compared to Sb.
	In the band structure of \ce{Cs3Sb} the valence band maximum is found at the high-symmetry point $X$, while the conduction band minimum is at $\Gamma$, making the system an indirect band-gap semiconductor.~\cite{cocc+19sr,sass-cocc21es, cocc-sass21micromachines}	
	\ce{Cs2Te} exhibits analogous features, except for a 0.5~eV larger value of the band gap amounting to 2.14~eV according to our SCAN calculations.~\cite{sass-cocc21es}
	The smallest direct band gap is located at the $\Gamma$-point having a value of 2.55~eV.	
	
	Turning to the band structures of the trigonal and hexagonal phases (Fig.~\ref{fig:bands}~c-d), the valence band maximum is located at the origin of the Brillouin zone, as for the orthorhombic structure (Fig.~\ref{fig:bands}~a).
	The primitive cell of \ce{Cs18Te9} hosts a total of 27 atoms and, thus, the band structure includes a large number of states (Fig.~\ref{fig:bands}~c).
	Specifically, the different inequivalent Te-sites result in a further stretched out set of valence bands spanning an energy range of about 1~eV and resulting in a slightly smaller band gap of 1.96~eV compared to the other structures.
	A similar energy range is spanned by the valence states of the hexagonal \ce{Cs4Te2} crystal (see Fig.~\ref{fig:bands}~d).
	However, in this case, bands are generally more dispersive and the band gap is similar in size to the ones of the orthorhombic and cubic structures, see Table~\ref{tab:Cs2Te_phases}.
	
As a final remark, it is worth noting that all the band structures displayed in Fig.~\ref{fig:bands} not only exhibit similar band-gap values but also the conduction-band minimum separated by approximately 1~eV from the rest of the conduction region.
Transitions to the lowest unoccupied band are weak but optically allowed while transitions to higher states give rise to a much broader and intenser absorption, as shown for \ce{Cs2Te} in recent work.~\cite{cocc-sass21micromachines}
These features are consistent with experimental data attributing large values of quantum efficiency to Cs-Te photocathodes for photon energies larger than 3.5~eV,~\cite{gaow+19prab,kong+95jap,sert+00nimpra,prat+15prab} in contrast with the much lower yield observed at frequencies below that threshold.~\cite{kong+95jap,sert+00nimpra}
	
	\section{Summary and conclusions}\label{sec:conclusions}
	
	In summary, we have presented a new computational workflow developed upon the existing AiiDA infrastructure to investigate the structural stability and the electronic properties of photocathode materials via high-throughput DFT calculations with the SCAN functional.
	An efficient data-mining routine integrated in the workflow enables users to take entries from open material databases as input structures.
	The applicability of the implemented approach is demonstrated in the study of the compositional phase space of Cs-Te binary systems. 
	Our results have revealed that the structures with the lowest formation energies have 2:1 relative content of Cs and Te atoms, respectively, in agreement with experimental findings. 
	Both metallic and semiconducting materials are identified among the stable compounds. 
	The band-gap computed with SCAN are systematically larger than those obtained from PBE and stored in the databases, thereby confirming that the usage of the meta-GGA functional enables alleviating the severe underestimation of this quantity by semi-local DFT without affecting the overall computational costs. 
	In the second part of our analysis, we focus on the electronic properties of selected stable phases, inspecting the pDOS of various structures at varying Cs:Te ratio. 
	Noting that Te $p$-orbitals (Cs $s$-orbitals) dominate the uppermost valence (lowermost conduction) region, we found a clear correlation between the relative Cs content and the electronic structure of the system: Te (Cs) prevalence leads towards a semiconducting (metallic) behavior, with an evident discontinuity of this trend at the 2:1 Cs:Te compositional content.
	To better understand this peculiar behavior of the phases with 2:1 ratio, which also correspond to the experimentally most relevant systems, we focus on the corresponding structures in our dataset, sorting them into 4 groups and analyzing the band structure calculated for one representative of each family.
	All phases with 2:1 Cs:Te composition show very similar band gaps ranging from approximately 2~eV to 2.25~eV, although differences appear in light of the varying stoichiometries and crystal structures.
	
	In conclusion, this study demonstrates the viability of high-throughput computational schemes based on advanced approximations of DFT xc functionals. 
	For materials like the cesium-tellurides where SCAN has been proven to predict the electronic properties in excellent agreement with calculations employing hybrid functionals or even based on the $GW$ method,~\cite{sass-cocc21es} this approximation can be efficiently integrated into an automated workflow and deliver results that are relevant to advance the corresponding research area. 
In particular, gaining insight into the stability and the electronic structure of candidate materials for photocathode applications can serve as a starting point to complement experimental research,  currently still based on inefficient and expensive trial-and-error procedures. Especially, recent advances towards single-crystal growth methods can further enhance this development by ensuring a more controlled experimental environment.~\cite{parz+21arxiv}
	Taking DFT results as input for the three-step model for photoemission has been recently proven as a viable method to predict operational properties of photocathode materials prior or in parallel to their experiment growth and characterization.~\cite{anto+20prb,nang+21prb}
To this end, dedicated studies on surface facets~\cite{schier-master} enabling the determination of relevant parameters such as work functions and electron affinities are urgently needed.
	Adopting for this purpose an efficient and accurate high-throughput scheme, such as the one presented in this work, will enable exploring a much larger pool of structures and compositions. 
	Last but certainly not least, the presented computational scheme is ready to be interfaced with molecular dynamics routines to access the thermodynamic properties of the systems, as well as with algorithms for crystal structure prediction to enlarge the pool of candidate structures beyond those stored in databases or generated manually. 
	Finally, the future integration of machine learning processes, for which the workflow is ready, will make the developed infrastructure ready for the new paradigm of materials science artificial intelligence.

		\section*{Supplementary Material}
We report the comparison between the projected density of states of \ce{Cs2Te} computed with the supercell method and by sampling directly the first Brillouin zone.~\cite{sass-cocc21es}
	
	\section*{Data Availability}

	The data produced in this study are openly available in Zenodo at http://doi.org/10.5281/zenodo.5779129, reference number 5779129.

	\begin{acknowledgments}
		We are grateful to Sven Lederer for inspiring discussions on Cs-Te photocathodes and to Timo Reents for exchanges on the computational workflow.
		We thank Jürg Hutter for the permission to use the GTH-pseudopotentials specifically parameterized for the SCAN functional.
		This work is funded by the German Federal Ministry of Education and Research (Professorinnenprogramm III) as well as from the Lower Saxony State (Professorinnen f\"ur Niedersachsen).
		Computational resources provided by the North-German Supercomputing Alliance (HLRN), projects bep00084 and nic00069, and by the HPC cluster CARL at the University of Oldenburg, funded by the DFG (project number INST 184/157-1 FUGG) and by the Ministry of Science and Culture of the Lower Saxony State.
		
	\end{acknowledgments}
	

\begin{thebibliography}{83}%
\makeatletter
\providecommand \@ifxundefined [1]{%
 \@ifx{#1\undefined}
}%
\providecommand \@ifnum [1]{%
 \ifnum #1\expandafter \@firstoftwo
 \else \expandafter \@secondoftwo
 \fi
}%
\providecommand \@ifx [1]{%
 \ifx #1\expandafter \@firstoftwo
 \else \expandafter \@secondoftwo
 \fi
}%
\providecommand \natexlab [1]{#1}%
\providecommand \enquote  [1]{``#1''}%
\providecommand \bibnamefont  [1]{#1}%
\providecommand \bibfnamefont [1]{#1}%
\providecommand \citenamefont [1]{#1}%
\providecommand \href@noop [0]{\@secondoftwo}%
\providecommand \href [0]{\begingroup \@sanitize@url \@href}%
\providecommand \@href[1]{\@@startlink{#1}\@@href}%
\providecommand \@@href[1]{\endgroup#1\@@endlink}%
\providecommand \@sanitize@url [0]{\catcode `\\12\catcode `\$12\catcode
  `\&12\catcode `\#12\catcode `\^12\catcode `\_12\catcode `\%12\relax}%
\providecommand \@@startlink[1]{}%
\providecommand \@@endlink[0]{}%
\providecommand \url  [0]{\begingroup\@sanitize@url \@url }%
\providecommand \@url [1]{\endgroup\@href {#1}{\urlprefix }}%
\providecommand \urlprefix  [0]{URL }%
\providecommand \Eprint [0]{\href }%
\providecommand \doibase [0]{https://doi.org/}%
\providecommand \selectlanguage [0]{\@gobble}%
\providecommand \bibinfo  [0]{\@secondoftwo}%
\providecommand \bibfield  [0]{\@secondoftwo}%
\providecommand \translation [1]{[#1]}%
\providecommand \BibitemOpen [0]{}%
\providecommand \bibitemStop [0]{}%
\providecommand \bibitemNoStop [0]{.\EOS\space}%
\providecommand \EOS [0]{\spacefactor3000\relax}%
\providecommand \BibitemShut  [1]{\csname bibitem#1\endcsname}%
\let\auto@bib@innerbib\@empty
\bibitem [{\citenamefont {Dowell}\ \emph {et~al.}(2010)\citenamefont {Dowell},
  \citenamefont {Bazarov}, \citenamefont {Dunham}, \citenamefont {Harkay},
  \citenamefont {Hernandez-Garcia}, \citenamefont {Legg}, \citenamefont
  {Padmore}, \citenamefont {Rao}, \citenamefont {Smedley},\ and\ \citenamefont
  {Wan}}]{dowe+10nimpra}%
  \BibitemOpen
  \bibfield  {author} {\bibinfo {author} {\bibfnamefont {D.}~\bibnamefont
  {Dowell}}, \bibinfo {author} {\bibfnamefont {I.}~\bibnamefont {Bazarov}},
  \bibinfo {author} {\bibfnamefont {B.}~\bibnamefont {Dunham}}, \bibinfo
  {author} {\bibfnamefont {K.}~\bibnamefont {Harkay}}, \bibinfo {author}
  {\bibfnamefont {C.}~\bibnamefont {Hernandez-Garcia}}, \bibinfo {author}
  {\bibfnamefont {R.}~\bibnamefont {Legg}}, \bibinfo {author} {\bibfnamefont
  {H.}~\bibnamefont {Padmore}}, \bibinfo {author} {\bibfnamefont
  {T.}~\bibnamefont {Rao}}, \bibinfo {author} {\bibfnamefont {J.}~\bibnamefont
  {Smedley}},\ and\ \bibinfo {author} {\bibfnamefont {W.}~\bibnamefont {Wan}},\
  }\bibfield  {title} {\enquote {\bibinfo {title} {Cathode r\&d for future
  light sources},}\ }\href@noop {} {\bibfield  {journal} {\bibinfo  {journal}
  {Nucl.~Instrum.~Methods~Phys.~Res.~A}\ }\textbf {\bibinfo {volume} {622}},\
  \bibinfo {pages} {685--697} (\bibinfo {year} {2010})}\BibitemShut {NoStop}%
\bibitem [{\citenamefont {Musumeci}\ \emph {et~al.}(2010)\citenamefont
  {Musumeci}, \citenamefont {Moody}, \citenamefont {Scoby}, \citenamefont
  {Gutierrez}, \citenamefont {Bender},\ and\ \citenamefont
  {Wilcox}}]{musu+10rsi}%
  \BibitemOpen
  \bibfield  {author} {\bibinfo {author} {\bibfnamefont {P.}~\bibnamefont
  {Musumeci}}, \bibinfo {author} {\bibfnamefont {J.}~\bibnamefont {Moody}},
  \bibinfo {author} {\bibfnamefont {C.}~\bibnamefont {Scoby}}, \bibinfo
  {author} {\bibfnamefont {M.}~\bibnamefont {Gutierrez}}, \bibinfo {author}
  {\bibfnamefont {H.}~\bibnamefont {Bender}},\ and\ \bibinfo {author}
  {\bibfnamefont {N.}~\bibnamefont {Wilcox}},\ }\bibfield  {title} {\enquote
  {\bibinfo {title} {High quality single shot diffraction patterns using
  ultrashort megaelectron volt electron beams from a radio frequency
  photoinjector},}\ }\href@noop {} {\bibfield  {journal} {\bibinfo  {journal}
  {Rev.~Sci.~Instrum.~}\ }\textbf {\bibinfo {volume} {81}},\ \bibinfo {pages}
  {013306} (\bibinfo {year} {2010})}\BibitemShut {NoStop}%
\bibitem [{\citenamefont {Musumeci}\ \emph {et~al.}(2018)\citenamefont
  {Musumeci}, \citenamefont {Navarro}, \citenamefont {Rosenzweig},
  \citenamefont {Cultrera}, \citenamefont {Bazarov}, \citenamefont {Maxson},
  \citenamefont {Karkare},\ and\ \citenamefont {Padmore}}]{musu+18nimpra}%
  \BibitemOpen
  \bibfield  {author} {\bibinfo {author} {\bibfnamefont {P.}~\bibnamefont
  {Musumeci}}, \bibinfo {author} {\bibfnamefont {J.~G.}\ \bibnamefont
  {Navarro}}, \bibinfo {author} {\bibfnamefont {J.}~\bibnamefont {Rosenzweig}},
  \bibinfo {author} {\bibfnamefont {L.}~\bibnamefont {Cultrera}}, \bibinfo
  {author} {\bibfnamefont {I.}~\bibnamefont {Bazarov}}, \bibinfo {author}
  {\bibfnamefont {J.}~\bibnamefont {Maxson}}, \bibinfo {author} {\bibfnamefont
  {S.}~\bibnamefont {Karkare}},\ and\ \bibinfo {author} {\bibfnamefont
  {H.}~\bibnamefont {Padmore}},\ }\bibfield  {title} {\enquote {\bibinfo
  {title} {Advances in bright electron sources},}\ }\href@noop {} {\bibfield
  {journal} {\bibinfo  {journal} {Nucl.~Instrum.~Methods~Phys.~Res.~A}\ }
  (\bibinfo {year} {2018})}\BibitemShut {NoStop}%
\bibitem [{\citenamefont {Boucher}\ \emph {et~al.}(2009)\citenamefont
  {Boucher}, \citenamefont {Frigola}, \citenamefont {Murokh}, \citenamefont
  {Ruelas}, \citenamefont {Jovanovic}, \citenamefont {Rosenzweig},\ and\
  \citenamefont {Travish}}]{bouc+09nimpra}%
  \BibitemOpen
  \bibfield  {author} {\bibinfo {author} {\bibfnamefont {S.}~\bibnamefont
  {Boucher}}, \bibinfo {author} {\bibfnamefont {P.}~\bibnamefont {Frigola}},
  \bibinfo {author} {\bibfnamefont {A.}~\bibnamefont {Murokh}}, \bibinfo
  {author} {\bibfnamefont {M.}~\bibnamefont {Ruelas}}, \bibinfo {author}
  {\bibfnamefont {I.}~\bibnamefont {Jovanovic}}, \bibinfo {author}
  {\bibfnamefont {J.}~\bibnamefont {Rosenzweig}},\ and\ \bibinfo {author}
  {\bibfnamefont {G.}~\bibnamefont {Travish}},\ }\bibfield  {title} {\enquote
  {\bibinfo {title} {Inverse compton scattering gamma ray source},}\
  }\href@noop {} {\bibfield  {journal} {\bibinfo  {journal}
  {Nucl.~Instrum.~Methods~Phys.~Res.~A}\ }\textbf {\bibinfo {volume} {608}},\
  \bibinfo {pages} {S54--S56} (\bibinfo {year} {2009})}\BibitemShut {NoStop}%
\bibitem [{\citenamefont {Chen}, \citenamefont {Zagorodnov},\ and\
  \citenamefont {Dohlus}(2020)}]{chen+20prab}%
  \BibitemOpen
  \bibfield  {author} {\bibinfo {author} {\bibfnamefont {Y.}~\bibnamefont
  {Chen}}, \bibinfo {author} {\bibfnamefont {I.}~\bibnamefont {Zagorodnov}},\
  and\ \bibinfo {author} {\bibfnamefont {M.}~\bibnamefont {Dohlus}},\
  }\bibfield  {title} {\enquote {\bibinfo {title} {Beam dynamics of realistic
  bunches at the injector section of the european x-ray free-electron laser},}\
  }\href@noop {} {\bibfield  {journal} {\bibinfo  {journal}
  {Phys.~Rev.~ST~Accel.~Beams}\ }\textbf {\bibinfo {volume} {23}},\ \bibinfo
  {pages} {044201} (\bibinfo {year} {2020})}\BibitemShut {NoStop}%
\bibitem [{\citenamefont {McConnell}(2021)}]{mcco21book}%
  \BibitemOpen
  \bibfield  {author} {\bibinfo {author} {\bibfnamefont {M.~L.}\ \bibnamefont
  {McConnell}},\ }\bibfield  {title} {\enquote {\bibinfo {title} {Scintillation
  detectors for x-ray and $\gamma$-ray astronomy},}\ }in\ \href@noop {} {\emph
  {\bibinfo {booktitle} {The WSPC Handbook of Astronomical Instrumentation:
  Volume 5: Gamma-Ray and Multimessenger Astronomical Instrumentation}}}\
  (\bibinfo  {publisher} {World Scientific},\ \bibinfo {year} {2021})\ pp.\
  \bibinfo {pages} {27--50}\BibitemShut {NoStop}%
\bibitem [{\citenamefont {Cocchi}\ and\ \citenamefont
  {Sa{\ss}nick}(2021)}]{cocc-sass21micromachines}%
  \BibitemOpen
  \bibfield  {author} {\bibinfo {author} {\bibfnamefont {C.}~\bibnamefont
  {Cocchi}}\ and\ \bibinfo {author} {\bibfnamefont {H.-D.}\ \bibnamefont
  {Sa{\ss}nick}},\ }\bibfield  {title} {\enquote {\bibinfo {title} {Ab initio
  quantum--mechanical predictions of semiconducting photocathode materials},}\
  }\href@noop {} {\bibfield  {journal} {\bibinfo  {journal} {Micromachines}\
  }\textbf {\bibinfo {volume} {12}},\ \bibinfo {pages} {1002} (\bibinfo {year}
  {2021})}\BibitemShut {NoStop}%
\bibitem [{\citenamefont {Antoniuk}\ \emph {et~al.}(2021)\citenamefont
  {Antoniuk}, \citenamefont {Schindler}, \citenamefont {Schroeder},
  \citenamefont {Dunham}, \citenamefont {Pianetta}, \citenamefont {Vecchione},\
  and\ \citenamefont {Reed}}]{anto+21am}%
  \BibitemOpen
  \bibfield  {author} {\bibinfo {author} {\bibfnamefont {E.~R.}\ \bibnamefont
  {Antoniuk}}, \bibinfo {author} {\bibfnamefont {P.}~\bibnamefont {Schindler}},
  \bibinfo {author} {\bibfnamefont {W.~A.}\ \bibnamefont {Schroeder}}, \bibinfo
  {author} {\bibfnamefont {B.}~\bibnamefont {Dunham}}, \bibinfo {author}
  {\bibfnamefont {P.}~\bibnamefont {Pianetta}}, \bibinfo {author}
  {\bibfnamefont {T.}~\bibnamefont {Vecchione}},\ and\ \bibinfo {author}
  {\bibfnamefont {E.~J.}\ \bibnamefont {Reed}},\ }\bibfield  {title} {\enquote
  {\bibinfo {title} {Novel ultrabright and air-stable photocathodes discovered
  from machine learning and density functional theory driven screening},}\
  }\href {https://doi.org/10.1002/adma.202104081} {\bibfield  {journal}
  {\bibinfo  {journal} {Adv.~Mater.~}\ }\textbf {\bibinfo {volume} {n/a}},\
  \bibinfo {pages} {2104081} (\bibinfo {year} {2021})},\ \Eprint
  {https://arxiv.org/abs/https://onlinelibrary.wiley.com/doi/pdf/10.1002/adma.202104081}
  {https://onlinelibrary.wiley.com/doi/pdf/10.1002/adma.202104081} \BibitemShut
  {NoStop}%
\bibitem [{\citenamefont {Louie}\ \emph {et~al.}(2021)\citenamefont {Louie},
  \citenamefont {Chan}, \citenamefont {Felipe}, \citenamefont {Li},\ and\
  \citenamefont {Qiu}}]{loui+21natm}%
  \BibitemOpen
  \bibfield  {author} {\bibinfo {author} {\bibfnamefont {S.~G.}\ \bibnamefont
  {Louie}}, \bibinfo {author} {\bibfnamefont {Y.-H.}\ \bibnamefont {Chan}},
  \bibinfo {author} {\bibfnamefont {H.}~\bibnamefont {Felipe}}, \bibinfo
  {author} {\bibfnamefont {Z.}~\bibnamefont {Li}},\ and\ \bibinfo {author}
  {\bibfnamefont {D.~Y.}\ \bibnamefont {Qiu}},\ }\bibfield  {title} {\enquote
  {\bibinfo {title} {Discovering and understanding materials through
  computation},}\ }\href@noop {} {\bibfield  {journal} {\bibinfo  {journal}
  {Nature Mater.}\ }\textbf {\bibinfo {volume} {20}},\ \bibinfo {pages}
  {728--735} (\bibinfo {year} {2021})}\BibitemShut {NoStop}%
\bibitem [{\citenamefont {Marzari}, \citenamefont {Ferretti},\ and\
  \citenamefont {Wolverton}(2021)}]{marz+21natm}%
  \BibitemOpen
  \bibfield  {author} {\bibinfo {author} {\bibfnamefont {N.}~\bibnamefont
  {Marzari}}, \bibinfo {author} {\bibfnamefont {A.}~\bibnamefont {Ferretti}},\
  and\ \bibinfo {author} {\bibfnamefont {C.}~\bibnamefont {Wolverton}},\
  }\bibfield  {title} {\enquote {\bibinfo {title} {Electronic-structure methods
  for materials design},}\ }\href@noop {} {\bibfield  {journal} {\bibinfo
  {journal} {Nature Mater.}\ }\textbf {\bibinfo {volume} {20}},\ \bibinfo
  {pages} {736--749} (\bibinfo {year} {2021})}\BibitemShut {NoStop}%
\bibitem [{\citenamefont {Curtarolo}\ \emph {et~al.}(2012)\citenamefont
  {Curtarolo}, \citenamefont {Setyawan}, \citenamefont {Hart}, \citenamefont
  {Jahnatek}, \citenamefont {Chepulskii}, \citenamefont {Taylor}, \citenamefont
  {Wang}, \citenamefont {Xue}, \citenamefont {Yang}, \citenamefont {Levy},
  \citenamefont {Mehl}, \citenamefont {Stokes}, \citenamefont {Demchenko},\
  and\ \citenamefont {Morgan}}]{curt+12cms}%
  \BibitemOpen
  \bibfield  {author} {\bibinfo {author} {\bibfnamefont {S.}~\bibnamefont
  {Curtarolo}}, \bibinfo {author} {\bibfnamefont {W.}~\bibnamefont {Setyawan}},
  \bibinfo {author} {\bibfnamefont {G.~L.}\ \bibnamefont {Hart}}, \bibinfo
  {author} {\bibfnamefont {M.}~\bibnamefont {Jahnatek}}, \bibinfo {author}
  {\bibfnamefont {R.~V.}\ \bibnamefont {Chepulskii}}, \bibinfo {author}
  {\bibfnamefont {R.~H.}\ \bibnamefont {Taylor}}, \bibinfo {author}
  {\bibfnamefont {S.}~\bibnamefont {Wang}}, \bibinfo {author} {\bibfnamefont
  {J.}~\bibnamefont {Xue}}, \bibinfo {author} {\bibfnamefont {K.}~\bibnamefont
  {Yang}}, \bibinfo {author} {\bibfnamefont {O.}~\bibnamefont {Levy}}, \bibinfo
  {author} {\bibfnamefont {M.~J.}\ \bibnamefont {Mehl}}, \bibinfo {author}
  {\bibfnamefont {H.~T.}\ \bibnamefont {Stokes}}, \bibinfo {author}
  {\bibfnamefont {D.~O.}\ \bibnamefont {Demchenko}},\ and\ \bibinfo {author}
  {\bibfnamefont {D.}~\bibnamefont {Morgan}},\ }\bibfield  {title} {\enquote
  {\bibinfo {title} {Aflow: An automatic framework for high-throughput
  materials discovery},}\ }\href
  {https://doi.org/https://doi.org/10.1016/j.commatsci.2012.02.005} {\bibfield
  {journal} {\bibinfo  {journal} {Comp.~Mater.~Sci.~}\ }\textbf {\bibinfo
  {volume} {58}},\ \bibinfo {pages} {218--226} (\bibinfo {year}
  {2012})}\BibitemShut {NoStop}%
\bibitem [{\citenamefont {Jain}\ \emph {et~al.}(2013)\citenamefont {Jain},
  \citenamefont {Ong}, \citenamefont {Hautier}, \citenamefont {Chen},
  \citenamefont {Richards}, \citenamefont {Dacek}, \citenamefont {Cholia},
  \citenamefont {Gunter}, \citenamefont {Skinner}, \citenamefont {Ceder},\ and\
  \citenamefont {Persson}}]{jain+13aplm}%
  \BibitemOpen
  \bibfield  {author} {\bibinfo {author} {\bibfnamefont {A.}~\bibnamefont
  {Jain}}, \bibinfo {author} {\bibfnamefont {S.~P.}\ \bibnamefont {Ong}},
  \bibinfo {author} {\bibfnamefont {G.}~\bibnamefont {Hautier}}, \bibinfo
  {author} {\bibfnamefont {W.}~\bibnamefont {Chen}}, \bibinfo {author}
  {\bibfnamefont {W.~D.}\ \bibnamefont {Richards}}, \bibinfo {author}
  {\bibfnamefont {S.}~\bibnamefont {Dacek}}, \bibinfo {author} {\bibfnamefont
  {S.}~\bibnamefont {Cholia}}, \bibinfo {author} {\bibfnamefont
  {D.}~\bibnamefont {Gunter}}, \bibinfo {author} {\bibfnamefont
  {D.}~\bibnamefont {Skinner}}, \bibinfo {author} {\bibfnamefont
  {G.}~\bibnamefont {Ceder}},\ and\ \bibinfo {author} {\bibfnamefont {K.~A.}\
  \bibnamefont {Persson}},\ }\bibfield  {title} {\enquote {\bibinfo {title}
  {Commentary: The materials project: A materials genome approach to
  accelerating materials innovation},}\ }\href
  {https://doi.org/10.1063/1.4812323} {\bibfield  {journal} {\bibinfo
  {journal} {APL~Mater.}\ }\textbf {\bibinfo {volume} {1}},\ \bibinfo {pages}
  {011002} (\bibinfo {year} {2013})},\ \Eprint
  {https://arxiv.org/abs/https://doi.org/10.1063/1.4812323}
  {https://doi.org/10.1063/1.4812323} \BibitemShut {NoStop}%
\bibitem [{\citenamefont {Saal}\ \emph {et~al.}(2013)\citenamefont {Saal},
  \citenamefont {Kirklin}, \citenamefont {Aykol}, \citenamefont {Meredig},\
  and\ \citenamefont {Wolverton}}]{saal+13jom}%
  \BibitemOpen
  \bibfield  {author} {\bibinfo {author} {\bibfnamefont {J.~E.}\ \bibnamefont
  {Saal}}, \bibinfo {author} {\bibfnamefont {S.}~\bibnamefont {Kirklin}},
  \bibinfo {author} {\bibfnamefont {M.}~\bibnamefont {Aykol}}, \bibinfo
  {author} {\bibfnamefont {B.}~\bibnamefont {Meredig}},\ and\ \bibinfo {author}
  {\bibfnamefont {C.}~\bibnamefont {Wolverton}},\ }\bibfield  {title} {\enquote
  {\bibinfo {title} {Materials design and discovery with high-throughput
  density functional theory: The open quantum materials database (oqmd)},}\
  }\href {https://doi.org/10.1007/s11837-013-0755-4} {\bibfield  {journal}
  {\bibinfo  {journal} {JOM}\ }\textbf {\bibinfo {volume} {65}},\ \bibinfo
  {pages} {1501--1509} (\bibinfo {year} {2013})}\BibitemShut {NoStop}%
\bibitem [{\citenamefont {Draxl}\ and\ \citenamefont
  {Scheffler}(2019)}]{drax-sche19jpm}%
  \BibitemOpen
  \bibfield  {author} {\bibinfo {author} {\bibfnamefont {C.}~\bibnamefont
  {Draxl}}\ and\ \bibinfo {author} {\bibfnamefont {M.}~\bibnamefont
  {Scheffler}},\ }\bibfield  {title} {\enquote {\bibinfo {title} {The nomad
  laboratory: from data sharing to artificial intelligence},}\ }\href@noop {}
  {\bibfield  {journal} {\bibinfo  {journal} {J.~Phys.~Mater.}\ }\textbf
  {\bibinfo {volume} {2}},\ \bibinfo {pages} {036001} (\bibinfo {year}
  {2019})}\BibitemShut {NoStop}%
\bibitem [{\citenamefont {Talirz}\ \emph {et~al.}(2020)\citenamefont {Talirz},
  \citenamefont {Kumbhar}, \citenamefont {Passaro}, \citenamefont {Yakutovich},
  \citenamefont {Granata}, \citenamefont {Gargiulo}, \citenamefont {Borelli},
  \citenamefont {Uhrin}, \citenamefont {Huber}, \citenamefont {Zoupanos},
  \citenamefont {Adorf}, \citenamefont {Andersen}, \citenamefont {Sch{\"u}tt},
  \citenamefont {Pignedoli}, \citenamefont {Passerone}, \citenamefont
  {VandeVondele}, \citenamefont {Schulthess}, \citenamefont {Smit},
  \citenamefont {Pizzi},\ and\ \citenamefont {Marzari}}]{tali+20sd}%
  \BibitemOpen
  \bibfield  {author} {\bibinfo {author} {\bibfnamefont {L.}~\bibnamefont
  {Talirz}}, \bibinfo {author} {\bibfnamefont {S.}~\bibnamefont {Kumbhar}},
  \bibinfo {author} {\bibfnamefont {E.}~\bibnamefont {Passaro}}, \bibinfo
  {author} {\bibfnamefont {A.~V.}\ \bibnamefont {Yakutovich}}, \bibinfo
  {author} {\bibfnamefont {V.}~\bibnamefont {Granata}}, \bibinfo {author}
  {\bibfnamefont {F.}~\bibnamefont {Gargiulo}}, \bibinfo {author}
  {\bibfnamefont {M.}~\bibnamefont {Borelli}}, \bibinfo {author} {\bibfnamefont
  {M.}~\bibnamefont {Uhrin}}, \bibinfo {author} {\bibfnamefont {S.~P.}\
  \bibnamefont {Huber}}, \bibinfo {author} {\bibfnamefont {S.}~\bibnamefont
  {Zoupanos}}, \bibinfo {author} {\bibfnamefont {C.~S.}\ \bibnamefont {Adorf}},
  \bibinfo {author} {\bibfnamefont {C.~W.}\ \bibnamefont {Andersen}}, \bibinfo
  {author} {\bibfnamefont {O.}~\bibnamefont {Sch{\"u}tt}}, \bibinfo {author}
  {\bibfnamefont {C.~A.}\ \bibnamefont {Pignedoli}}, \bibinfo {author}
  {\bibfnamefont {D.}~\bibnamefont {Passerone}}, \bibinfo {author}
  {\bibfnamefont {J.}~\bibnamefont {VandeVondele}}, \bibinfo {author}
  {\bibfnamefont {T.~C.}\ \bibnamefont {Schulthess}}, \bibinfo {author}
  {\bibfnamefont {B.}~\bibnamefont {Smit}}, \bibinfo {author} {\bibfnamefont
  {G.}~\bibnamefont {Pizzi}},\ and\ \bibinfo {author} {\bibfnamefont
  {N.}~\bibnamefont {Marzari}},\ }\bibfield  {title} {\enquote {\bibinfo
  {title} {Materials cloud, a platform for open computational science},}\
  }\href {https://doi.org/10.1038/s41597-020-00637-5} {\bibfield  {journal}
  {\bibinfo  {journal} {Scientific Data}\ }\textbf {\bibinfo {volume} {7}},\
  \bibinfo {pages} {299} (\bibinfo {year} {2020})}\BibitemShut {NoStop}%
\bibitem [{\citenamefont {Gossett}\ \emph {et~al.}(2018)\citenamefont
  {Gossett}, \citenamefont {Toher}, \citenamefont {Oses}, \citenamefont
  {Isayev}, \citenamefont {Legrain}, \citenamefont {Rose}, \citenamefont
  {Zurek}, \citenamefont {Carrete}, \citenamefont {Mingo}, \citenamefont
  {Tropsha}, \citenamefont {Mingo}, \citenamefont {Tropsha},\ and\
  \citenamefont {Curtarolo}}]{goss+18cms}%
  \BibitemOpen
  \bibfield  {author} {\bibinfo {author} {\bibfnamefont {E.}~\bibnamefont
  {Gossett}}, \bibinfo {author} {\bibfnamefont {C.}~\bibnamefont {Toher}},
  \bibinfo {author} {\bibfnamefont {C.}~\bibnamefont {Oses}}, \bibinfo {author}
  {\bibfnamefont {O.}~\bibnamefont {Isayev}}, \bibinfo {author} {\bibfnamefont
  {F.}~\bibnamefont {Legrain}}, \bibinfo {author} {\bibfnamefont
  {F.}~\bibnamefont {Rose}}, \bibinfo {author} {\bibfnamefont {E.}~\bibnamefont
  {Zurek}}, \bibinfo {author} {\bibfnamefont {J.}~\bibnamefont {Carrete}},
  \bibinfo {author} {\bibfnamefont {N.}~\bibnamefont {Mingo}}, \bibinfo
  {author} {\bibfnamefont {A.}~\bibnamefont {Tropsha}}, \bibinfo {author}
  {\bibfnamefont {N.}~\bibnamefont {Mingo}}, \bibinfo {author} {\bibfnamefont
  {A.}~\bibnamefont {Tropsha}},\ and\ \bibinfo {author} {\bibfnamefont
  {S.}~\bibnamefont {Curtarolo}},\ }\bibfield  {title} {\enquote {\bibinfo
  {title} {Aflow-ml: A restful api for machine-learning predictions of
  materials properties},}\ }\href@noop {} {\bibfield  {journal} {\bibinfo
  {journal} {Comp.~Mater.~Sci.~}\ }\textbf {\bibinfo {volume} {152}},\ \bibinfo
  {pages} {134--145} (\bibinfo {year} {2018})}\BibitemShut {NoStop}%
\bibitem [{\citenamefont {Pizzi}, \citenamefont {Togo},\ and\ \citenamefont
  {Kozinsky}(2018)}]{pizzi+18mrs}%
  \BibitemOpen
  \bibfield  {author} {\bibinfo {author} {\bibfnamefont {G.}~\bibnamefont
  {Pizzi}}, \bibinfo {author} {\bibfnamefont {A.}~\bibnamefont {Togo}},\ and\
  \bibinfo {author} {\bibfnamefont {B.}~\bibnamefont {Kozinsky}},\ }\bibfield
  {title} {\enquote {\bibinfo {title} {Provenance, workflows, and
  crystallographic tools in materials science: Aiida, spglib, and seekpath},}\
  }\href {https://doi.org/10.1557/mrs.2018.203} {\bibfield  {journal} {\bibinfo
   {journal} {MRS~Bull.}\ }\textbf {\bibinfo {volume} {43}},\ \bibinfo {pages}
  {696–702} (\bibinfo {year} {2018})}\BibitemShut {NoStop}%
\bibitem [{\citenamefont {Yakutovich}\ \emph {et~al.}(2021)\citenamefont
  {Yakutovich}, \citenamefont {Eimre}, \citenamefont {Schütt}, \citenamefont
  {Talirz}, \citenamefont {Adorf}, \citenamefont {Andersen}, \citenamefont
  {Ditler}, \citenamefont {Du}, \citenamefont {Passerone}, \citenamefont
  {Smit}, \citenamefont {Marzari}, \citenamefont {Pizzi},\ and\ \citenamefont
  {Pignedoli}}]{yaku+21cms}%
  \BibitemOpen
  \bibfield  {author} {\bibinfo {author} {\bibfnamefont {A.~V.}\ \bibnamefont
  {Yakutovich}}, \bibinfo {author} {\bibfnamefont {K.}~\bibnamefont {Eimre}},
  \bibinfo {author} {\bibfnamefont {O.}~\bibnamefont {Schütt}}, \bibinfo
  {author} {\bibfnamefont {L.}~\bibnamefont {Talirz}}, \bibinfo {author}
  {\bibfnamefont {C.~S.}\ \bibnamefont {Adorf}}, \bibinfo {author}
  {\bibfnamefont {C.~W.}\ \bibnamefont {Andersen}}, \bibinfo {author}
  {\bibfnamefont {E.}~\bibnamefont {Ditler}}, \bibinfo {author} {\bibfnamefont
  {D.}~\bibnamefont {Du}}, \bibinfo {author} {\bibfnamefont {D.}~\bibnamefont
  {Passerone}}, \bibinfo {author} {\bibfnamefont {B.}~\bibnamefont {Smit}},
  \bibinfo {author} {\bibfnamefont {N.}~\bibnamefont {Marzari}}, \bibinfo
  {author} {\bibfnamefont {G.}~\bibnamefont {Pizzi}},\ and\ \bibinfo {author}
  {\bibfnamefont {C.~A.}\ \bibnamefont {Pignedoli}},\ }\bibfield  {title}
  {\enquote {\bibinfo {title} {Aiidalab – an ecosystem for developing,
  executing, and sharing scientific workflows},}\ }\href
  {https://doi.org/https://doi.org/10.1016/j.commatsci.2020.110165} {\bibfield
  {journal} {\bibinfo  {journal} {Comp.~Mater.~Sci.~}\ }\textbf {\bibinfo
  {volume} {188}},\ \bibinfo {pages} {110165} (\bibinfo {year}
  {2021})}\BibitemShut {NoStop}%
\bibitem [{\citenamefont {Schleder}\ \emph {et~al.}(2019)\citenamefont
  {Schleder}, \citenamefont {Padilha}, \citenamefont {Acosta}, \citenamefont
  {Costa},\ and\ \citenamefont {Fazzio}}]{schl+19jpm}%
  \BibitemOpen
  \bibfield  {author} {\bibinfo {author} {\bibfnamefont {G.~R.}\ \bibnamefont
  {Schleder}}, \bibinfo {author} {\bibfnamefont {A.~C.}\ \bibnamefont
  {Padilha}}, \bibinfo {author} {\bibfnamefont {C.~M.}\ \bibnamefont {Acosta}},
  \bibinfo {author} {\bibfnamefont {M.}~\bibnamefont {Costa}},\ and\ \bibinfo
  {author} {\bibfnamefont {A.}~\bibnamefont {Fazzio}},\ }\bibfield  {title}
  {\enquote {\bibinfo {title} {From dft to machine learning: recent approaches
  to materials science--a review},}\ }\href@noop {} {\bibfield  {journal}
  {\bibinfo  {journal} {J.~Phys.~Mater.}\ }\textbf {\bibinfo {volume} {2}},\
  \bibinfo {pages} {032001} (\bibinfo {year} {2019})}\BibitemShut {NoStop}%
\bibitem [{\citenamefont {Chibani}\ and\ \citenamefont
  {Coudert}(2020)}]{chib-coud20aplm}%
  \BibitemOpen
  \bibfield  {author} {\bibinfo {author} {\bibfnamefont {S.}~\bibnamefont
  {Chibani}}\ and\ \bibinfo {author} {\bibfnamefont {F.-X.}\ \bibnamefont
  {Coudert}},\ }\bibfield  {title} {\enquote {\bibinfo {title} {Machine
  learning approaches for the prediction of materials properties},}\
  }\href@noop {} {\bibfield  {journal} {\bibinfo  {journal} {APL~Mater.}\
  }\textbf {\bibinfo {volume} {8}},\ \bibinfo {pages} {080701} (\bibinfo {year}
  {2020})}\BibitemShut {NoStop}%
\bibitem [{\citenamefont {Cheng}\ \emph {et~al.}(2020)\citenamefont {Cheng},
  \citenamefont {Zhu}, \citenamefont {Wang}, \citenamefont {Zhou},
  \citenamefont {Elliott},\ and\ \citenamefont {Sun}}]{chen+20cms}%
  \BibitemOpen
  \bibfield  {author} {\bibinfo {author} {\bibfnamefont {Y.}~\bibnamefont
  {Cheng}}, \bibinfo {author} {\bibfnamefont {L.}~\bibnamefont {Zhu}}, \bibinfo
  {author} {\bibfnamefont {G.}~\bibnamefont {Wang}}, \bibinfo {author}
  {\bibfnamefont {J.}~\bibnamefont {Zhou}}, \bibinfo {author} {\bibfnamefont
  {S.~R.}\ \bibnamefont {Elliott}},\ and\ \bibinfo {author} {\bibfnamefont
  {Z.}~\bibnamefont {Sun}},\ }\bibfield  {title} {\enquote {\bibinfo {title}
  {Vacancy formation energy and its connection with bonding environment in
  solid: A high-throughput calculation and machine learning study},}\
  }\href@noop {} {\bibfield  {journal} {\bibinfo  {journal}
  {Comp.~Mater.~Sci.~}\ }\textbf {\bibinfo {volume} {183}},\ \bibinfo {pages}
  {109803} (\bibinfo {year} {2020})}\BibitemShut {NoStop}%
\bibitem [{\citenamefont {Sahni}\ \emph {et~al.}(2020)\citenamefont {Sahni},
  \citenamefont {Vikram}, \citenamefont {Kangsabanik},\ and\ \citenamefont
  {Alam}}]{sahn+20jpcl}%
  \BibitemOpen
  \bibfield  {author} {\bibinfo {author} {\bibfnamefont {B.}~\bibnamefont
  {Sahni}}, \bibinfo {author} {\bibnamefont {Vikram}}, \bibinfo {author}
  {\bibfnamefont {J.}~\bibnamefont {Kangsabanik}},\ and\ \bibinfo {author}
  {\bibfnamefont {A.}~\bibnamefont {Alam}},\ }\bibfield  {title} {\enquote
  {\bibinfo {title} {Reliable prediction of new quantum materials for
  topological and renewable-energy applications: A high-throughput
  screening},}\ }\href@noop {} {\bibfield  {journal} {\bibinfo  {journal}
  {J.~Phys.~Chem.~Lett.}\ }\textbf {\bibinfo {volume} {11}},\ \bibinfo {pages}
  {6364--6372} (\bibinfo {year} {2020})}\BibitemShut {NoStop}%
\bibitem [{\citenamefont {Ran}\ \emph {et~al.}(2021)\citenamefont {Ran},
  \citenamefont {Sun}, \citenamefont {Qiu}, \citenamefont {Song}, \citenamefont
  {Chen},\ and\ \citenamefont {Liu}}]{ran+21jpcl}%
  \BibitemOpen
  \bibfield  {author} {\bibinfo {author} {\bibfnamefont {N.}~\bibnamefont
  {Ran}}, \bibinfo {author} {\bibfnamefont {B.}~\bibnamefont {Sun}}, \bibinfo
  {author} {\bibfnamefont {W.}~\bibnamefont {Qiu}}, \bibinfo {author}
  {\bibfnamefont {E.}~\bibnamefont {Song}}, \bibinfo {author} {\bibfnamefont
  {T.}~\bibnamefont {Chen}},\ and\ \bibinfo {author} {\bibfnamefont
  {J.}~\bibnamefont {Liu}},\ }\bibfield  {title} {\enquote {\bibinfo {title}
  {Identifying metallic transition-metal dichalcogenides for hydrogen evolution
  through multilevel high-throughput calculations and machine learning},}\
  }\href@noop {} {\bibfield  {journal} {\bibinfo  {journal}
  {J.~Phys.~Chem.~Lett.}\ }\textbf {\bibinfo {volume} {12}},\ \bibinfo {pages}
  {2102--2111} (\bibinfo {year} {2021})}\BibitemShut {NoStop}%
\bibitem [{\citenamefont {Dahliah}\ \emph {et~al.}(2021)\citenamefont
  {Dahliah}, \citenamefont {Brunin}, \citenamefont {George}, \citenamefont
  {Ha}, \citenamefont {Rignanese},\ and\ \citenamefont {Hautier}}]{dahl+21ees}%
  \BibitemOpen
  \bibfield  {author} {\bibinfo {author} {\bibfnamefont {D.}~\bibnamefont
  {Dahliah}}, \bibinfo {author} {\bibfnamefont {G.}~\bibnamefont {Brunin}},
  \bibinfo {author} {\bibfnamefont {J.}~\bibnamefont {George}}, \bibinfo
  {author} {\bibfnamefont {V.-A.}\ \bibnamefont {Ha}}, \bibinfo {author}
  {\bibfnamefont {G.-M.}\ \bibnamefont {Rignanese}},\ and\ \bibinfo {author}
  {\bibfnamefont {G.}~\bibnamefont {Hautier}},\ }\bibfield  {title} {\enquote
  {\bibinfo {title} {High-throughput computational search for high carrier
  lifetime, defect-tolerant solar absorbers},}\ }\href@noop {} {\bibfield
  {journal} {\bibinfo  {journal} {Energy~Environ.~Sci.}\ }\textbf {\bibinfo
  {volume} {14}},\ \bibinfo {pages} {5057--5073} (\bibinfo {year}
  {2021})}\BibitemShut {NoStop}%
\bibitem [{\citenamefont {Onida}, \citenamefont {Reining},\ and\ \citenamefont
  {Rubio}(2002)}]{onid+02rmp}%
  \BibitemOpen
  \bibfield  {author} {\bibinfo {author} {\bibfnamefont {G.}~\bibnamefont
  {Onida}}, \bibinfo {author} {\bibfnamefont {L.}~\bibnamefont {Reining}},\
  and\ \bibinfo {author} {\bibfnamefont {A.}~\bibnamefont {Rubio}},\ }\bibfield
   {title} {\enquote {\bibinfo {title} {Electronic excitations:
  density-functional versus many-body green’s-function approaches},}\
  }\href@noop {} {\bibfield  {journal} {\bibinfo  {journal} {Rev.~Mod.~Phys.~}\
  }\textbf {\bibinfo {volume} {74}},\ \bibinfo {pages} {601} (\bibinfo {year}
  {2002})}\BibitemShut {NoStop}%
\bibitem [{\citenamefont {Lee}, \citenamefont {Yang},\ and\ \citenamefont
  {Parr}(1988)}]{b3lyp}%
  \BibitemOpen
  \bibfield  {author} {\bibinfo {author} {\bibfnamefont {C.}~\bibnamefont
  {Lee}}, \bibinfo {author} {\bibfnamefont {W.}~\bibnamefont {Yang}},\ and\
  \bibinfo {author} {\bibfnamefont {R.~G.}\ \bibnamefont {Parr}},\ }\bibfield
  {title} {\enquote {\bibinfo {title} {Development of the colle-salvetti
  correlation-energy formula into a functional of the electron density},}\
  }\href {https://doi.org/10.1103/PhysRevB.37.785} {\bibfield  {journal}
  {\bibinfo  {journal} {Phys.~Rev.~B}\ }\textbf {\bibinfo {volume} {37}},\
  \bibinfo {pages} {785--789} (\bibinfo {year} {1988})}\BibitemShut {NoStop}%
\bibitem [{\citenamefont {Adamo}\ and\ \citenamefont
  {Barone}(1999)}]{adam-baro99jcp}%
  \BibitemOpen
  \bibfield  {author} {\bibinfo {author} {\bibfnamefont {C.}~\bibnamefont
  {Adamo}}\ and\ \bibinfo {author} {\bibfnamefont {V.}~\bibnamefont {Barone}},\
  }\bibfield  {title} {\enquote {\bibinfo {title} {Toward reliable density
  functional methods without adjustable parameters: The pbe0 model},}\
  }\href@noop {} {\bibfield  {journal} {\bibinfo  {journal} {J.~Chem.~Phys.~}\
  }\textbf {\bibinfo {volume} {110}},\ \bibinfo {pages} {6158--6170} (\bibinfo
  {year} {1999})}\BibitemShut {NoStop}%
\bibitem [{\citenamefont {Heyd}, \citenamefont {Scuseria},\ and\ \citenamefont
  {Ernzerhof}(2003)}]{hse03}%
  \BibitemOpen
  \bibfield  {author} {\bibinfo {author} {\bibfnamefont {J.}~\bibnamefont
  {Heyd}}, \bibinfo {author} {\bibfnamefont {G.~E.}\ \bibnamefont {Scuseria}},\
  and\ \bibinfo {author} {\bibfnamefont {M.}~\bibnamefont {Ernzerhof}},\
  }\bibfield  {title} {\enquote {\bibinfo {title} {Hybrid functionals based on
  a screened coulomb potential},}\ }\href@noop {} {\bibfield  {journal}
  {\bibinfo  {journal} {J.~Chem.~Phys.~}\ }\textbf {\bibinfo {volume} {118}},\
  \bibinfo {pages} {8207--8215} (\bibinfo {year} {2003})}\BibitemShut {NoStop}%
\bibitem [{\citenamefont {Garza}\ and\ \citenamefont
  {Scuseria}(2016)}]{garz-scus16jpcl}%
  \BibitemOpen
  \bibfield  {author} {\bibinfo {author} {\bibfnamefont {A.~J.}\ \bibnamefont
  {Garza}}\ and\ \bibinfo {author} {\bibfnamefont {G.~E.}\ \bibnamefont
  {Scuseria}},\ }\bibfield  {title} {\enquote {\bibinfo {title} {Predicting
  band gaps with hybrid density functionals},}\ }\href@noop {} {\bibfield
  {journal} {\bibinfo  {journal} {J.~Phys.~Chem.~Lett.}\ }\textbf {\bibinfo
  {volume} {7}},\ \bibinfo {pages} {4165--4170} (\bibinfo {year}
  {2016})}\BibitemShut {NoStop}%
\bibitem [{\citenamefont {Borlido}\ \emph {et~al.}(2019)\citenamefont
  {Borlido}, \citenamefont {Aull}, \citenamefont {Huran}, \citenamefont {Tran},
  \citenamefont {Marques},\ and\ \citenamefont {Botti}}]{borl+19jctc}%
  \BibitemOpen
  \bibfield  {author} {\bibinfo {author} {\bibfnamefont {P.}~\bibnamefont
  {Borlido}}, \bibinfo {author} {\bibfnamefont {T.}~\bibnamefont {Aull}},
  \bibinfo {author} {\bibfnamefont {A.~W.}\ \bibnamefont {Huran}}, \bibinfo
  {author} {\bibfnamefont {F.}~\bibnamefont {Tran}}, \bibinfo {author}
  {\bibfnamefont {M.~A.}\ \bibnamefont {Marques}},\ and\ \bibinfo {author}
  {\bibfnamefont {S.}~\bibnamefont {Botti}},\ }\bibfield  {title} {\enquote
  {\bibinfo {title} {Large-scale benchmark of exchange--correlation functionals
  for the determination of electronic band gaps of solids},}\ }\href@noop {}
  {\bibfield  {journal} {\bibinfo  {journal} {J.~Chem.~Theory.~Comput.~}\
  }\textbf {\bibinfo {volume} {15}},\ \bibinfo {pages} {5069--5079} (\bibinfo
  {year} {2019})}\BibitemShut {NoStop}%
\bibitem [{\citenamefont {Burke}(2012)}]{burk12jcp}%
  \BibitemOpen
  \bibfield  {author} {\bibinfo {author} {\bibfnamefont {K.}~\bibnamefont
  {Burke}},\ }\bibfield  {title} {\enquote {\bibinfo {title} {Perspective on
  density functional theory},}\ }\href@noop {} {\bibfield  {journal} {\bibinfo
  {journal} {J.~Chem.~Phys.~}\ }\textbf {\bibinfo {volume} {136}},\ \bibinfo
  {pages} {150901} (\bibinfo {year} {2012})}\BibitemShut {NoStop}%
\bibitem [{\citenamefont {Sun}, \citenamefont {Ruzsinszky},\ and\ \citenamefont
  {Perdew}(2015)}]{sun+15prl}%
  \BibitemOpen
  \bibfield  {author} {\bibinfo {author} {\bibfnamefont {J.}~\bibnamefont
  {Sun}}, \bibinfo {author} {\bibfnamefont {A.}~\bibnamefont {Ruzsinszky}},\
  and\ \bibinfo {author} {\bibfnamefont {J.~P.}\ \bibnamefont {Perdew}},\
  }\bibfield  {title} {\enquote {\bibinfo {title} {Strongly constrained and
  appropriately normed semilocal density functional},}\ }\href
  {https://doi.org/10.1103/PhysRevLett.115.036402} {\bibfield  {journal}
  {\bibinfo  {journal} {Phys. Rev. Lett.}\ }\textbf {\bibinfo {volume} {115}},\
  \bibinfo {pages} {036402} (\bibinfo {year} {2015})}\BibitemShut {NoStop}%
\bibitem [{\citenamefont {Jana}, \citenamefont {Sharma},\ and\ \citenamefont
  {Samal}(2018)}]{jana+18jcp}%
  \BibitemOpen
  \bibfield  {author} {\bibinfo {author} {\bibfnamefont {S.}~\bibnamefont
  {Jana}}, \bibinfo {author} {\bibfnamefont {K.}~\bibnamefont {Sharma}},\ and\
  \bibinfo {author} {\bibfnamefont {P.}~\bibnamefont {Samal}},\ }\bibfield
  {title} {\enquote {\bibinfo {title} {Assessing the performance of the recent
  meta-gga density functionals for describing the lattice constants, bulk
  moduli, and cohesive energies of alkali, alkaline-earth, and transition
  metals},}\ }\href@noop {} {\bibfield  {journal} {\bibinfo  {journal}
  {J.~Chem.~Phys.~}\ }\textbf {\bibinfo {volume} {149}},\ \bibinfo {pages}
  {164703} (\bibinfo {year} {2018})}\BibitemShut {NoStop}%
\bibitem [{\citenamefont {Sa{\ss}nick}\ and\ \citenamefont
  {Cocchi}(2021)}]{sass-cocc21es}%
  \BibitemOpen
  \bibfield  {author} {\bibinfo {author} {\bibfnamefont {H.-D.}\ \bibnamefont
  {Sa{\ss}nick}}\ and\ \bibinfo {author} {\bibfnamefont {C.}~\bibnamefont
  {Cocchi}},\ }\bibfield  {title} {\enquote {\bibinfo {title} {Electronic
  structure of cesium-based photocathode materials from density functional
  theory: performance of {PBE}, {SCAN}, and {HSE}06 functionals},}\ }\href
  {https://doi.org/10.1088/2516-1075/abfb08} {\bibfield  {journal} {\bibinfo
  {journal} {Electr.~Struct.}\ }\textbf {\bibinfo {volume} {3}},\ \bibinfo
  {pages} {027001} (\bibinfo {year} {2021})}\BibitemShut {NoStop}%
\bibitem [{\citenamefont {Pisch}\ \emph {et~al.}(2020)\citenamefont {Pisch},
  \citenamefont {Pasturel}, \citenamefont {Deffrennes}, \citenamefont
  {Dezellus}, \citenamefont {Benigni},\ and\ \citenamefont
  {Mikaelian}}]{pisc+20cms}%
  \BibitemOpen
  \bibfield  {author} {\bibinfo {author} {\bibfnamefont {A.}~\bibnamefont
  {Pisch}}, \bibinfo {author} {\bibfnamefont {A.}~\bibnamefont {Pasturel}},
  \bibinfo {author} {\bibfnamefont {G.}~\bibnamefont {Deffrennes}}, \bibinfo
  {author} {\bibfnamefont {O.}~\bibnamefont {Dezellus}}, \bibinfo {author}
  {\bibfnamefont {P.}~\bibnamefont {Benigni}},\ and\ \bibinfo {author}
  {\bibfnamefont {G.}~\bibnamefont {Mikaelian}},\ }\bibfield  {title} {\enquote
  {\bibinfo {title} {Investigation of the thermodynamic properties of al4c3: A
  combined dft and dsc study},}\ }\href
  {https://doi.org/https://doi.org/10.1016/j.commatsci.2019.109100} {\bibfield
  {journal} {\bibinfo  {journal} {Comp.~Mater.~Sci.~}\ }\textbf {\bibinfo
  {volume} {171}},\ \bibinfo {pages} {109100} (\bibinfo {year}
  {2020})}\BibitemShut {NoStop}%
\bibitem [{\citenamefont {Yao}\ and\ \citenamefont
  {Kanai}(2017)}]{yao-kana17jcp}%
  \BibitemOpen
  \bibfield  {author} {\bibinfo {author} {\bibfnamefont {Y.}~\bibnamefont
  {Yao}}\ and\ \bibinfo {author} {\bibfnamefont {Y.}~\bibnamefont {Kanai}},\
  }\bibfield  {title} {\enquote {\bibinfo {title} {Plane-wave pseudopotential
  implementation and performance of scan meta-gga exchange-correlation
  functional for extended systems},}\ }\href@noop {} {\bibfield  {journal}
  {\bibinfo  {journal} {J.~Chem.~Phys.~}\ }\textbf {\bibinfo {volume} {146}},\
  \bibinfo {pages} {224105} (\bibinfo {year} {2017})}\BibitemShut {NoStop}%
\bibitem [{\citenamefont {Kong}\ \emph {et~al.}(1995)\citenamefont {Kong},
  \citenamefont {Kinross-Wright}, \citenamefont {Nguyen},\ and\ \citenamefont
  {Sheffield}}]{kong+95jap}%
  \BibitemOpen
  \bibfield  {author} {\bibinfo {author} {\bibfnamefont {S.}~\bibnamefont
  {Kong}}, \bibinfo {author} {\bibfnamefont {J.}~\bibnamefont
  {Kinross-Wright}}, \bibinfo {author} {\bibfnamefont {D.}~\bibnamefont
  {Nguyen}},\ and\ \bibinfo {author} {\bibfnamefont {R.}~\bibnamefont
  {Sheffield}},\ }\bibfield  {title} {\enquote {\bibinfo {title} {Cesium
  telluride photocathodes},}\ }\href@noop {} {\bibfield  {journal} {\bibinfo
  {journal} {J.~Appl.~Phys.~}\ }\textbf {\bibinfo {volume} {77}},\ \bibinfo
  {pages} {6031--6038} (\bibinfo {year} {1995})}\BibitemShut {NoStop}%
\bibitem [{\citenamefont {Kühne}\ \emph {et~al.}(2020)\citenamefont {Kühne},
  \citenamefont {Iannuzzi}, \citenamefont {Del~Ben}, \citenamefont {Rybkin},
  \citenamefont {Seewald}, \citenamefont {Stein}, \citenamefont {Laino},
  \citenamefont {Khaliullin}, \citenamefont {Schütt}, \citenamefont
  {Schiffmann}, \citenamefont {Golze}, \citenamefont {Wilhelm}, \citenamefont
  {Chulkov}, \citenamefont {Bani-Hashemian}, \citenamefont {Weber},
  \citenamefont {Borštnik}, \citenamefont {Taillefumier}, \citenamefont
  {Jakobovits}, \citenamefont {Lazzaro}, \citenamefont {Pabst}, \citenamefont
  {Müller}, \citenamefont {Schade}, \citenamefont {Guidon}, \citenamefont
  {Andermatt}, \citenamefont {Holmberg}, \citenamefont {Schenter},
  \citenamefont {Hehn}, \citenamefont {Bussy}, \citenamefont {Belleflamme},
  \citenamefont {Tabacchi}, \citenamefont {Glöß}, \citenamefont {Lass},
  \citenamefont {Bethune}, \citenamefont {Mundy}, \citenamefont {Plessl},
  \citenamefont {Watkins}, \citenamefont {VandeVondele}, \citenamefont
  {Krack},\ and\ \citenamefont {Hutter}}]{kueh+20jcp}%
  \BibitemOpen
  \bibfield  {author} {\bibinfo {author} {\bibfnamefont {T.~D.}\ \bibnamefont
  {Kühne}}, \bibinfo {author} {\bibfnamefont {M.}~\bibnamefont {Iannuzzi}},
  \bibinfo {author} {\bibfnamefont {M.}~\bibnamefont {Del~Ben}}, \bibinfo
  {author} {\bibfnamefont {V.~V.}\ \bibnamefont {Rybkin}}, \bibinfo {author}
  {\bibfnamefont {P.}~\bibnamefont {Seewald}}, \bibinfo {author} {\bibfnamefont
  {F.}~\bibnamefont {Stein}}, \bibinfo {author} {\bibfnamefont
  {T.}~\bibnamefont {Laino}}, \bibinfo {author} {\bibfnamefont {R.~Z.}\
  \bibnamefont {Khaliullin}}, \bibinfo {author} {\bibfnamefont
  {O.}~\bibnamefont {Schütt}}, \bibinfo {author} {\bibfnamefont
  {F.}~\bibnamefont {Schiffmann}}, \bibinfo {author} {\bibfnamefont
  {D.}~\bibnamefont {Golze}}, \bibinfo {author} {\bibfnamefont
  {J.}~\bibnamefont {Wilhelm}}, \bibinfo {author} {\bibfnamefont
  {S.}~\bibnamefont {Chulkov}}, \bibinfo {author} {\bibfnamefont {M.~H.}\
  \bibnamefont {Bani-Hashemian}}, \bibinfo {author} {\bibfnamefont
  {V.}~\bibnamefont {Weber}}, \bibinfo {author} {\bibfnamefont
  {U.}~\bibnamefont {Borštnik}}, \bibinfo {author} {\bibfnamefont
  {M.}~\bibnamefont {Taillefumier}}, \bibinfo {author} {\bibfnamefont {A.~S.}\
  \bibnamefont {Jakobovits}}, \bibinfo {author} {\bibfnamefont
  {A.}~\bibnamefont {Lazzaro}}, \bibinfo {author} {\bibfnamefont
  {H.}~\bibnamefont {Pabst}}, \bibinfo {author} {\bibfnamefont
  {T.}~\bibnamefont {Müller}}, \bibinfo {author} {\bibfnamefont
  {R.}~\bibnamefont {Schade}}, \bibinfo {author} {\bibfnamefont
  {M.}~\bibnamefont {Guidon}}, \bibinfo {author} {\bibfnamefont
  {S.}~\bibnamefont {Andermatt}}, \bibinfo {author} {\bibfnamefont
  {N.}~\bibnamefont {Holmberg}}, \bibinfo {author} {\bibfnamefont {G.~K.}\
  \bibnamefont {Schenter}}, \bibinfo {author} {\bibfnamefont {A.}~\bibnamefont
  {Hehn}}, \bibinfo {author} {\bibfnamefont {A.}~\bibnamefont {Bussy}},
  \bibinfo {author} {\bibfnamefont {F.}~\bibnamefont {Belleflamme}}, \bibinfo
  {author} {\bibfnamefont {G.}~\bibnamefont {Tabacchi}}, \bibinfo {author}
  {\bibfnamefont {A.}~\bibnamefont {Glöß}}, \bibinfo {author} {\bibfnamefont
  {M.}~\bibnamefont {Lass}}, \bibinfo {author} {\bibfnamefont {I.}~\bibnamefont
  {Bethune}}, \bibinfo {author} {\bibfnamefont {C.~J.}\ \bibnamefont {Mundy}},
  \bibinfo {author} {\bibfnamefont {C.}~\bibnamefont {Plessl}}, \bibinfo
  {author} {\bibfnamefont {M.}~\bibnamefont {Watkins}}, \bibinfo {author}
  {\bibfnamefont {J.}~\bibnamefont {VandeVondele}}, \bibinfo {author}
  {\bibfnamefont {M.}~\bibnamefont {Krack}},\ and\ \bibinfo {author}
  {\bibfnamefont {J.}~\bibnamefont {Hutter}},\ }\bibfield  {title} {\enquote
  {\bibinfo {title} {Cp2k: An electronic structure and molecular dynamics
  software package - quickstep: Efficient and accurate electronic structure
  calculations},}\ }\href {https://doi.org/10.1063/5.0007045} {\bibfield
  {journal} {\bibinfo  {journal} {J.~Chem.~Phys.~}\ }\textbf {\bibinfo {volume}
  {152}},\ \bibinfo {pages} {194103} (\bibinfo {year} {2020})}\BibitemShut
  {NoStop}%
\bibitem [{\citenamefont {Berglund}\ and\ \citenamefont
  {Spicer}(1964)}]{berg-spic64pr}%
  \BibitemOpen
  \bibfield  {author} {\bibinfo {author} {\bibfnamefont {C.~N.}\ \bibnamefont
  {Berglund}}\ and\ \bibinfo {author} {\bibfnamefont {W.~E.}\ \bibnamefont
  {Spicer}},\ }\bibfield  {title} {\enquote {\bibinfo {title} {Photoemission
  studies of copper and silver: theory},}\ }\href@noop {} {\bibfield  {journal}
  {\bibinfo  {journal} {Phys.~Rev.~}\ }\textbf {\bibinfo {volume} {136}},\
  \bibinfo {pages} {A1030} (\bibinfo {year} {1964})}\BibitemShut {NoStop}%
\bibitem [{\citenamefont {Huber}\ \emph {et~al.}(2020)\citenamefont {Huber},
  \citenamefont {Zoupanos}, \citenamefont {Uhrin}, \citenamefont {Talirz},
  \citenamefont {Kahle}, \citenamefont {H{\"a}uselmann}, \citenamefont
  {Gresch}, \citenamefont {M{\"u}ller}, \citenamefont {Yakutovich},
  \citenamefont {Andersen}, \citenamefont {Ramirez}, \citenamefont {Adorf},
  \citenamefont {Gargiulo}, \citenamefont {Kumbhar}, \citenamefont {Passaro},
  \citenamefont {Johnston}, \citenamefont {Merkys}, \citenamefont {Cepellotti},
  \citenamefont {Mounet}, \citenamefont {Marzari}, \citenamefont {Kozinsky},\
  and\ \citenamefont {Pizzi}}]{hube+20sd}%
  \BibitemOpen
  \bibfield  {author} {\bibinfo {author} {\bibfnamefont {S.~P.}\ \bibnamefont
  {Huber}}, \bibinfo {author} {\bibfnamefont {S.}~\bibnamefont {Zoupanos}},
  \bibinfo {author} {\bibfnamefont {M.}~\bibnamefont {Uhrin}}, \bibinfo
  {author} {\bibfnamefont {L.}~\bibnamefont {Talirz}}, \bibinfo {author}
  {\bibfnamefont {L.}~\bibnamefont {Kahle}}, \bibinfo {author} {\bibfnamefont
  {R.}~\bibnamefont {H{\"a}uselmann}}, \bibinfo {author} {\bibfnamefont
  {D.}~\bibnamefont {Gresch}}, \bibinfo {author} {\bibfnamefont
  {T.}~\bibnamefont {M{\"u}ller}}, \bibinfo {author} {\bibfnamefont {A.~V.}\
  \bibnamefont {Yakutovich}}, \bibinfo {author} {\bibfnamefont {C.~W.}\
  \bibnamefont {Andersen}}, \bibinfo {author} {\bibfnamefont {F.~F.}\
  \bibnamefont {Ramirez}}, \bibinfo {author} {\bibfnamefont {C.~S.}\
  \bibnamefont {Adorf}}, \bibinfo {author} {\bibfnamefont {F.}~\bibnamefont
  {Gargiulo}}, \bibinfo {author} {\bibfnamefont {S.}~\bibnamefont {Kumbhar}},
  \bibinfo {author} {\bibfnamefont {E.}~\bibnamefont {Passaro}}, \bibinfo
  {author} {\bibfnamefont {C.}~\bibnamefont {Johnston}}, \bibinfo {author}
  {\bibfnamefont {A.}~\bibnamefont {Merkys}}, \bibinfo {author} {\bibfnamefont
  {A.}~\bibnamefont {Cepellotti}}, \bibinfo {author} {\bibfnamefont
  {N.}~\bibnamefont {Mounet}}, \bibinfo {author} {\bibfnamefont
  {N.}~\bibnamefont {Marzari}}, \bibinfo {author} {\bibfnamefont
  {B.}~\bibnamefont {Kozinsky}},\ and\ \bibinfo {author} {\bibfnamefont
  {G.}~\bibnamefont {Pizzi}},\ }\bibfield  {title} {\enquote {\bibinfo {title}
  {Aiida 1.0, a scalable computational infrastructure for automated
  reproducible workflows and data provenance},}\ }\href
  {https://doi.org/10.1038/s41597-020-00638-4} {\bibfield  {journal} {\bibinfo
  {journal} {Sci.~Data}\ }\textbf {\bibinfo {volume} {7}},\ \bibinfo {pages}
  {300} (\bibinfo {year} {2020})}\BibitemShut {NoStop}%
\bibitem [{\citenamefont {Uhrin}\ \emph {et~al.}(2021)\citenamefont {Uhrin},
  \citenamefont {Huber}, \citenamefont {Yu}, \citenamefont {Marzari},\ and\
  \citenamefont {Pizzi}}]{uhri+21cms}%
  \BibitemOpen
  \bibfield  {author} {\bibinfo {author} {\bibfnamefont {M.}~\bibnamefont
  {Uhrin}}, \bibinfo {author} {\bibfnamefont {S.~P.}\ \bibnamefont {Huber}},
  \bibinfo {author} {\bibfnamefont {J.}~\bibnamefont {Yu}}, \bibinfo {author}
  {\bibfnamefont {N.}~\bibnamefont {Marzari}},\ and\ \bibinfo {author}
  {\bibfnamefont {G.}~\bibnamefont {Pizzi}},\ }\bibfield  {title} {\enquote
  {\bibinfo {title} {Workflows in aiida: Engineering a high-throughput,
  event-based engine for robust and modular computational workflows},}\ }\href
  {https://doi.org/https://doi.org/10.1016/j.commatsci.2020.110086} {\bibfield
  {journal} {\bibinfo  {journal} {Comp.~Mater.~Sci.~}\ }\textbf {\bibinfo
  {volume} {187}},\ \bibinfo {pages} {110086} (\bibinfo {year}
  {2021})}\BibitemShut {NoStop}%
\bibitem [{aii(2021)}]{aiida_cp2k21github}%
  \BibitemOpen
  \href@noop {} {\enquote {\bibinfo {title} {aiida-cp2k},}\ }\bibinfo
  {howpublished} {\url{https://github.com/aiidateam/aiida-cp2k}} (\bibinfo
  {year} {2021})\BibitemShut {NoStop}%
\bibitem [{\citenamefont {Pizzi}\ \emph {et~al.}(2016)\citenamefont {Pizzi},
  \citenamefont {Cepellotti}, \citenamefont {Sabatini}, \citenamefont
  {Marzari},\ and\ \citenamefont {Kozinsky}}]{pizzi+16cms}%
  \BibitemOpen
  \bibfield  {author} {\bibinfo {author} {\bibfnamefont {G.}~\bibnamefont
  {Pizzi}}, \bibinfo {author} {\bibfnamefont {A.}~\bibnamefont {Cepellotti}},
  \bibinfo {author} {\bibfnamefont {R.}~\bibnamefont {Sabatini}}, \bibinfo
  {author} {\bibfnamefont {N.}~\bibnamefont {Marzari}},\ and\ \bibinfo {author}
  {\bibfnamefont {B.}~\bibnamefont {Kozinsky}},\ }\bibfield  {title} {\enquote
  {\bibinfo {title} {Aiida: automated interactive infrastructure and database
  for computational science},}\ }\href
  {https://doi.org/https://doi.org/10.1016/j.commatsci.2015.09.013} {\bibfield
  {journal} {\bibinfo  {journal} {Comp.~Mater.~Sci.~}\ }\textbf {\bibinfo
  {volume} {111}},\ \bibinfo {pages} {218--230} (\bibinfo {year}
  {2016})}\BibitemShut {NoStop}%
\bibitem [{\citenamefont {Hinuma}\ \emph {et~al.}(2017)\citenamefont {Hinuma},
  \citenamefont {Pizzi}, \citenamefont {Kumagai}, \citenamefont {Oba},\ and\
  \citenamefont {Tanaka}}]{hinu+17cms}%
  \BibitemOpen
  \bibfield  {author} {\bibinfo {author} {\bibfnamefont {Y.}~\bibnamefont
  {Hinuma}}, \bibinfo {author} {\bibfnamefont {G.}~\bibnamefont {Pizzi}},
  \bibinfo {author} {\bibfnamefont {Y.}~\bibnamefont {Kumagai}}, \bibinfo
  {author} {\bibfnamefont {F.}~\bibnamefont {Oba}},\ and\ \bibinfo {author}
  {\bibfnamefont {I.}~\bibnamefont {Tanaka}},\ }\bibfield  {title} {\enquote
  {\bibinfo {title} {Band structure diagram paths based on crystallography},}\
  }\href {https://doi.org/https://doi.org/10.1016/j.commatsci.2016.10.015}
  {\bibfield  {journal} {\bibinfo  {journal} {Comp.~Mater.~Sci.~}\ }\textbf
  {\bibinfo {volume} {128}},\ \bibinfo {pages} {140--184} (\bibinfo {year}
  {2017})}\BibitemShut {NoStop}%
\bibitem [{\citenamefont {Togo}\ and\ \citenamefont
  {Tanaka}(2018)}]{togo-tana18arxiv}%
  \BibitemOpen
  \bibfield  {author} {\bibinfo {author} {\bibfnamefont {A.}~\bibnamefont
  {Togo}}\ and\ \bibinfo {author} {\bibfnamefont {I.}~\bibnamefont {Tanaka}},\
  }\href@noop {} {\enquote {\bibinfo {title} {$\texttt{Spglib}$: a software
  library for crystal symmetry search},}\ } (\bibinfo {year} {2018}),\ \Eprint
  {https://arxiv.org/abs/1808.01590} {arXiv:1808.01590 [cond-mat.mtrl-sci]}
  \BibitemShut {NoStop}%
\bibitem [{\citenamefont {Kohn}\ and\ \citenamefont
  {Sham}(1965)}]{kohn-sham65pr}%
  \BibitemOpen
  \bibfield  {author} {\bibinfo {author} {\bibfnamefont {W.}~\bibnamefont
  {Kohn}}\ and\ \bibinfo {author} {\bibfnamefont {L.~J.}\ \bibnamefont
  {Sham}},\ }\bibfield  {title} {\enquote {\bibinfo {title} {Self-consistent
  equations including exchange and correlation effects},}\ }\href
  {https://doi.org/10.1103/PhysRev.140.A1133} {\bibfield  {journal} {\bibinfo
  {journal} {Phys.~Rev.~}\ }\textbf {\bibinfo {volume} {140}},\ \bibinfo
  {pages} {A1133--A1138} (\bibinfo {year} {1965})}\BibitemShut {NoStop}%
\bibitem [{\citenamefont {Baarman}\ and\ \citenamefont
  {VandeVondele}(2011)}]{baar-vand11jcp}%
  \BibitemOpen
  \bibfield  {author} {\bibinfo {author} {\bibfnamefont {K.}~\bibnamefont
  {Baarman}}\ and\ \bibinfo {author} {\bibfnamefont {J.}~\bibnamefont
  {VandeVondele}},\ }\bibfield  {title} {\enquote {\bibinfo {title} {A
  comparison of accelerators for direct energy minimization in electronic
  structure calculations},}\ }\href {https://doi.org/10.1063/1.3603445}
  {\bibfield  {journal} {\bibinfo  {journal} {J.~Chem.~Phys.~}\ }\textbf
  {\bibinfo {volume} {134}},\ \bibinfo {pages} {244104} (\bibinfo {year}
  {2011})},\ \Eprint {https://arxiv.org/abs/https://doi.org/10.1063/1.3603445}
  {https://doi.org/10.1063/1.3603445} \BibitemShut {NoStop}%
\bibitem [{\citenamefont {VandeVondele}\ and\ \citenamefont
  {Hutter}(2003)}]{vand-hutt03jcp}%
  \BibitemOpen
  \bibfield  {author} {\bibinfo {author} {\bibfnamefont {J.}~\bibnamefont
  {VandeVondele}}\ and\ \bibinfo {author} {\bibfnamefont {J.}~\bibnamefont
  {Hutter}},\ }\bibfield  {title} {\enquote {\bibinfo {title} {An efficient
  orbital transformation method for electronic structure calculations},}\
  }\href {https://doi.org/10.1063/1.1543154} {\bibfield  {journal} {\bibinfo
  {journal} {J.~Chem.~Phys.~}\ }\textbf {\bibinfo {volume} {118}},\ \bibinfo
  {pages} {4365--4369} (\bibinfo {year} {2003})},\ \Eprint
  {https://arxiv.org/abs/https://doi.org/10.1063/1.1543154}
  {https://doi.org/10.1063/1.1543154} \BibitemShut {NoStop}%
\bibitem [{\citenamefont {VandeVondele}\ \emph {et~al.}(2005)\citenamefont
  {VandeVondele}, \citenamefont {Krack}, \citenamefont {Mohamed}, \citenamefont
  {Parrinello}, \citenamefont {Chassaing},\ and\ \citenamefont
  {Hutter}}]{vand+05cpc}%
  \BibitemOpen
  \bibfield  {author} {\bibinfo {author} {\bibfnamefont {J.}~\bibnamefont
  {VandeVondele}}, \bibinfo {author} {\bibfnamefont {M.}~\bibnamefont {Krack}},
  \bibinfo {author} {\bibfnamefont {F.}~\bibnamefont {Mohamed}}, \bibinfo
  {author} {\bibfnamefont {M.}~\bibnamefont {Parrinello}}, \bibinfo {author}
  {\bibfnamefont {T.}~\bibnamefont {Chassaing}},\ and\ \bibinfo {author}
  {\bibfnamefont {J.}~\bibnamefont {Hutter}},\ }\bibfield  {title} {\enquote
  {\bibinfo {title} {Quickstep: Fast and accurate density functional
  calculations using a mixed gaussian and plane waves approach},}\ }\href
  {https://doi.org/https://doi.org/10.1016/j.cpc.2004.12.014} {\bibfield
  {journal} {\bibinfo  {journal} {Comput.~Phys.~Commun.~}\ }\textbf {\bibinfo
  {volume} {167}},\ \bibinfo {pages} {103--128} (\bibinfo {year}
  {2005})}\BibitemShut {NoStop}%
\bibitem [{\citenamefont {Goedecker}, \citenamefont {Teter},\ and\
  \citenamefont {Hutter}(1996)}]{goed+96prb}%
  \BibitemOpen
  \bibfield  {author} {\bibinfo {author} {\bibfnamefont {S.}~\bibnamefont
  {Goedecker}}, \bibinfo {author} {\bibfnamefont {M.}~\bibnamefont {Teter}},\
  and\ \bibinfo {author} {\bibfnamefont {J.}~\bibnamefont {Hutter}},\
  }\bibfield  {title} {\enquote {\bibinfo {title} {Separable dual-space
  gaussian pseudopotentials},}\ }\href
  {https://doi.org/10.1103/PhysRevB.54.1703} {\bibfield  {journal} {\bibinfo
  {journal} {Phys.~Rev.~B}\ }\textbf {\bibinfo {volume} {54}},\ \bibinfo
  {pages} {1703--1710} (\bibinfo {year} {1996})}\BibitemShut {NoStop}%
\bibitem [{\citenamefont {Hutter}(2021)}]{hutt21github}%
  \BibitemOpen
  \bibfield  {author} {\bibinfo {author} {\bibfnamefont {J.}~\bibnamefont
  {Hutter}},\ }\href@noop {} {\enquote {\bibinfo {title} {Gth},}\ }\bibinfo
  {howpublished} {\url{https://github.com/juerghutter/GTH/tree/master/SCAN}}
  (\bibinfo {year} {2021})\BibitemShut {NoStop}%
\bibitem [{\citenamefont {Powell}\ \emph {et~al.}(1973)\citenamefont {Powell},
  \citenamefont {Spicer}, \citenamefont {Fisher},\ and\ \citenamefont
  {Gregory}}]{powe+73prb}%
  \BibitemOpen
  \bibfield  {author} {\bibinfo {author} {\bibfnamefont {R.}~\bibnamefont
  {Powell}}, \bibinfo {author} {\bibfnamefont {W.}~\bibnamefont {Spicer}},
  \bibinfo {author} {\bibfnamefont {G.}~\bibnamefont {Fisher}},\ and\ \bibinfo
  {author} {\bibfnamefont {P.}~\bibnamefont {Gregory}},\ }\bibfield  {title}
  {\enquote {\bibinfo {title} {Photoemission studies of cesium telluride},}\
  }\href@noop {} {\bibfield  {journal} {\bibinfo  {journal} {Phys.~Rev.~B}\
  }\textbf {\bibinfo {volume} {8}},\ \bibinfo {pages} {3987} (\bibinfo {year}
  {1973})}\BibitemShut {NoStop}%
\bibitem [{\citenamefont {Gaowei}\ \emph {et~al.}(2019)\citenamefont {Gaowei},
  \citenamefont {Sinsheimer}, \citenamefont {Strom}, \citenamefont {Xie},
  \citenamefont {Cen}, \citenamefont {Walsh}, \citenamefont {Muller},\ and\
  \citenamefont {Smedley}}]{gaow+19prab}%
  \BibitemOpen
  \bibfield  {author} {\bibinfo {author} {\bibfnamefont {M.}~\bibnamefont
  {Gaowei}}, \bibinfo {author} {\bibfnamefont {J.}~\bibnamefont {Sinsheimer}},
  \bibinfo {author} {\bibfnamefont {D.}~\bibnamefont {Strom}}, \bibinfo
  {author} {\bibfnamefont {J.}~\bibnamefont {Xie}}, \bibinfo {author}
  {\bibfnamefont {J.}~\bibnamefont {Cen}}, \bibinfo {author} {\bibfnamefont
  {J.}~\bibnamefont {Walsh}}, \bibinfo {author} {\bibfnamefont
  {E.}~\bibnamefont {Muller}},\ and\ \bibinfo {author} {\bibfnamefont
  {J.}~\bibnamefont {Smedley}},\ }\bibfield  {title} {\enquote {\bibinfo
  {title} {Codeposition of ultrasmooth and high quantum efficiency cesium
  telluride photocathodes},}\ }\href@noop {} {\bibfield  {journal} {\bibinfo
  {journal} {Phys.~Rev.~ST~Accel.~Beams}\ }\textbf {\bibinfo {volume} {22}},\
  \bibinfo {pages} {073401} (\bibinfo {year} {2019})}\BibitemShut {NoStop}%
\bibitem [{\citenamefont {Terdik}\ \emph {et~al.}(2012)\citenamefont {Terdik},
  \citenamefont {N{\'e}meth}, \citenamefont {Harkay}, \citenamefont {Terry~Jr},
  \citenamefont {Spentzouris}, \citenamefont {Vel{\'a}zquez}, \citenamefont
  {Rosenberg},\ and\ \citenamefont {Srajer}}]{terd+12prb}%
  \BibitemOpen
  \bibfield  {author} {\bibinfo {author} {\bibfnamefont {J.~Z.}\ \bibnamefont
  {Terdik}}, \bibinfo {author} {\bibfnamefont {K.}~\bibnamefont {N{\'e}meth}},
  \bibinfo {author} {\bibfnamefont {K.~C.}\ \bibnamefont {Harkay}}, \bibinfo
  {author} {\bibfnamefont {J.~H.}\ \bibnamefont {Terry~Jr}}, \bibinfo {author}
  {\bibfnamefont {L.}~\bibnamefont {Spentzouris}}, \bibinfo {author}
  {\bibfnamefont {D.}~\bibnamefont {Vel{\'a}zquez}}, \bibinfo {author}
  {\bibfnamefont {R.}~\bibnamefont {Rosenberg}},\ and\ \bibinfo {author}
  {\bibfnamefont {G.}~\bibnamefont {Srajer}},\ }\bibfield  {title} {\enquote
  {\bibinfo {title} {Anomalous work function anisotropy in ternary
  acetylides},}\ }\href@noop {} {\bibfield  {journal} {\bibinfo  {journal}
  {Phys.~Rev.~B}\ }\textbf {\bibinfo {volume} {86}},\ \bibinfo {pages} {035142}
  (\bibinfo {year} {2012})}\BibitemShut {NoStop}%
\bibitem [{\citenamefont {Kalarasse}, \citenamefont {Bennecer},\ and\
  \citenamefont {Kalarasse}(2010)}]{kala+10jpcs}%
  \BibitemOpen
  \bibfield  {author} {\bibinfo {author} {\bibfnamefont {L.}~\bibnamefont
  {Kalarasse}}, \bibinfo {author} {\bibfnamefont {B.}~\bibnamefont
  {Bennecer}},\ and\ \bibinfo {author} {\bibfnamefont {F.}~\bibnamefont
  {Kalarasse}},\ }\bibfield  {title} {\enquote {\bibinfo {title} {Optical
  properties of the alkali antimonide semiconductors cs3sb, cs2ksb, csk2sb and
  k3sb},}\ }\href@noop {} {\bibfield  {journal} {\bibinfo  {journal}
  {J.~Phys.~Chem.~Solids}\ }\textbf {\bibinfo {volume} {71}},\ \bibinfo {pages}
  {314--322} (\bibinfo {year} {2010})}\BibitemShut {NoStop}%
\bibitem [{\citenamefont {Alay-e Abbas}\ and\ \citenamefont
  {Shaukat}(2011)}]{alay-shau11jms}%
  \BibitemOpen
  \bibfield  {author} {\bibinfo {author} {\bibfnamefont {S.}~\bibnamefont
  {Alay-e Abbas}}\ and\ \bibinfo {author} {\bibfnamefont {A.}~\bibnamefont
  {Shaukat}},\ }\bibfield  {title} {\enquote {\bibinfo {title} {Fp-lapw
  calculations of structural, electronic, and optical properties of alkali
  metal tellurides: M 2 te [m: Li, na, k and rb]},}\ }\href@noop {} {\bibfield
  {journal} {\bibinfo  {journal} {J.~Mol.~Struct.~}\ }\textbf {\bibinfo
  {volume} {46}},\ \bibinfo {pages} {1027--1037} (\bibinfo {year}
  {2011})}\BibitemShut {NoStop}%
\bibitem [{\citenamefont {Murtaza}\ \emph {et~al.}(2016)\citenamefont
  {Murtaza}, \citenamefont {Ullah}, \citenamefont {Ullah}, \citenamefont
  {Rani}, \citenamefont {Muzammil}, \citenamefont {Khenata}, \citenamefont
  {Ramay},\ and\ \citenamefont {Khan}}]{murt+16bms}%
  \BibitemOpen
  \bibfield  {author} {\bibinfo {author} {\bibfnamefont {G.}~\bibnamefont
  {Murtaza}}, \bibinfo {author} {\bibfnamefont {M.}~\bibnamefont {Ullah}},
  \bibinfo {author} {\bibfnamefont {N.}~\bibnamefont {Ullah}}, \bibinfo
  {author} {\bibfnamefont {M.}~\bibnamefont {Rani}}, \bibinfo {author}
  {\bibfnamefont {M.}~\bibnamefont {Muzammil}}, \bibinfo {author}
  {\bibfnamefont {R.}~\bibnamefont {Khenata}}, \bibinfo {author} {\bibfnamefont
  {S.~M.}\ \bibnamefont {Ramay}},\ and\ \bibinfo {author} {\bibfnamefont
  {U.}~\bibnamefont {Khan}},\ }\bibfield  {title} {\enquote {\bibinfo {title}
  {Structural, elastic, electronic and optical properties of bi-alkali
  antimonides},}\ }\href@noop {} {\bibfield  {journal} {\bibinfo  {journal}
  {Bull.~Mater.~Sci.}\ }\textbf {\bibinfo {volume} {39}},\ \bibinfo {pages}
  {1581--1591} (\bibinfo {year} {2016})}\BibitemShut {NoStop}%
\bibitem [{\citenamefont {Cocchi}\ \emph {et~al.}(2018)\citenamefont {Cocchi},
  \citenamefont {Mistry}, \citenamefont {Schmei{\ss}er}, \citenamefont
  {K{\"u}hn},\ and\ \citenamefont {Kamps}}]{cocc+18jpcm}%
  \BibitemOpen
  \bibfield  {author} {\bibinfo {author} {\bibfnamefont {C.}~\bibnamefont
  {Cocchi}}, \bibinfo {author} {\bibfnamefont {S.}~\bibnamefont {Mistry}},
  \bibinfo {author} {\bibfnamefont {M.}~\bibnamefont {Schmei{\ss}er}}, \bibinfo
  {author} {\bibfnamefont {J.}~\bibnamefont {K{\"u}hn}},\ and\ \bibinfo
  {author} {\bibfnamefont {T.}~\bibnamefont {Kamps}},\ }\bibfield  {title}
  {\enquote {\bibinfo {title} {First-principles many-body study of the
  electronic and optical properties of csk2sb, a semiconducting material for
  ultra-bright electron sources},}\ }\href@noop {} {\bibfield  {journal}
  {\bibinfo  {journal} {J.~Phys.~Condens.~Matter.~}\ }\textbf {\bibinfo
  {volume} {31}},\ \bibinfo {pages} {014002} (\bibinfo {year}
  {2018})}\BibitemShut {NoStop}%
\bibitem [{\citenamefont {Cocchi}\ \emph {et~al.}(2019)\citenamefont {Cocchi},
  \citenamefont {Mistry}, \citenamefont {Schmei{\ss}er}, \citenamefont
  {Amador}, \citenamefont {K{\"u}hn},\ and\ \citenamefont {Kamps}}]{cocc+19sr}%
  \BibitemOpen
  \bibfield  {author} {\bibinfo {author} {\bibfnamefont {C.}~\bibnamefont
  {Cocchi}}, \bibinfo {author} {\bibfnamefont {S.}~\bibnamefont {Mistry}},
  \bibinfo {author} {\bibfnamefont {M.}~\bibnamefont {Schmei{\ss}er}}, \bibinfo
  {author} {\bibfnamefont {R.}~\bibnamefont {Amador}}, \bibinfo {author}
  {\bibfnamefont {J.}~\bibnamefont {K{\"u}hn}},\ and\ \bibinfo {author}
  {\bibfnamefont {T.}~\bibnamefont {Kamps}},\ }\bibfield  {title} {\enquote
  {\bibinfo {title} {Electronic structure and core electron fingerprints of
  caesium-based multi-alkali antimonides for ultra-bright electron sources},}\
  }\href@noop {} {\bibfield  {journal} {\bibinfo  {journal} {Sci.~Rep.~}\
  }\textbf {\bibinfo {volume} {9}},\ \bibinfo {pages} {1--12} (\bibinfo {year}
  {2019})}\BibitemShut {NoStop}%
\bibitem [{\citenamefont {Cocchi}(2020)}]{cocc20pssrrl}%
  \BibitemOpen
  \bibfield  {author} {\bibinfo {author} {\bibfnamefont {C.}~\bibnamefont
  {Cocchi}},\ }\bibfield  {title} {\enquote {\bibinfo {title} {X-ray absorption
  fingerprints from cs atoms in cs3sb},}\ }\href
  {https://doi.org/10.1002/pssr.202000194} {\bibfield  {journal} {\bibinfo
  {journal} {Phys.~Status~Solidi~(RRL)}\ }\textbf {\bibinfo {volume} {14}},\
  \bibinfo {pages} {2000194} (\bibinfo {year} {2020})}\BibitemShut {NoStop}%
\bibitem [{\citenamefont {Amador}, \citenamefont {Sa{\ss}nick},\ and\
  \citenamefont {Cocchi}(2021)}]{amad+21jpcm}%
  \BibitemOpen
  \bibfield  {author} {\bibinfo {author} {\bibfnamefont {R.}~\bibnamefont
  {Amador}}, \bibinfo {author} {\bibfnamefont {H.-D.}\ \bibnamefont
  {Sa{\ss}nick}},\ and\ \bibinfo {author} {\bibfnamefont {C.}~\bibnamefont
  {Cocchi}},\ }\bibfield  {title} {\enquote {\bibinfo {title} {Electronic
  structure and optical properties of na2ksb and {NaK}2sb from first-principles
  many-body theory},}\ }\href {https://doi.org/10.1088/1361-648x/ac0e70}
  {\bibfield  {journal} {\bibinfo  {journal} {J.~Phys.~Condens.~Matter.~}\
  }\textbf {\bibinfo {volume} {33}},\ \bibinfo {pages} {365502} (\bibinfo
  {year} {2021})}\BibitemShut {NoStop}%
\bibitem [{\citenamefont {Schewe-Miller}\ and\ \citenamefont
  {B\"ottcher}(1991)}]{sche-boet91}%
  \BibitemOpen
  \bibfield  {author} {\bibinfo {author} {\bibfnamefont {I.}~\bibnamefont
  {Schewe-Miller}}\ and\ \bibinfo {author} {\bibfnamefont {P.}~\bibnamefont
  {B\"ottcher}},\ }\bibfield  {title} {\enquote {\bibinfo {title} {Synthesis
  and crystal structures of {K} $_{\textrm{5}}$ {Se} $_{\textrm{3}}$ , {Cs}
  $_{\textrm{5}}$ {Te} $_{\textrm{3}}$ and {Cs} $_{\textrm{2}}$ {Te}},}\ }\href
  {https://doi.org/10.1524/zkri.1991.196.1-4.137} {\bibfield  {journal}
  {\bibinfo  {journal} {Z.~Kristallogr.}\ }\textbf {\bibinfo {volume} {196}},\
  \bibinfo {pages} {137--151} (\bibinfo {year} {1991})}\BibitemShut {NoStop}%
\bibitem [{\citenamefont {Oganov}\ and\ \citenamefont
  {Valle}(2009)}]{ogan-vall09jcp}%
  \BibitemOpen
  \bibfield  {author} {\bibinfo {author} {\bibfnamefont {A.~R.}\ \bibnamefont
  {Oganov}}\ and\ \bibinfo {author} {\bibfnamefont {M.}~\bibnamefont {Valle}},\
  }\bibfield  {title} {\enquote {\bibinfo {title} {How to quantify energy
  landscapes of solids},}\ }\href {https://doi.org/10.1063/1.3079326}
  {\bibfield  {journal} {\bibinfo  {journal} {J.~Chem.~Phys.~}\ }\textbf
  {\bibinfo {volume} {130}},\ \bibinfo {pages} {104504} (\bibinfo {year}
  {2009})},\ \Eprint {https://arxiv.org/abs/https://doi.org/10.1063/1.3079326}
  {https://doi.org/10.1063/1.3079326} \BibitemShut {NoStop}%
\bibitem [{\citenamefont {Kresse}\ and\ \citenamefont
  {Furthm\"uller}(1996)}]{kres-furt96prb}%
  \BibitemOpen
  \bibfield  {author} {\bibinfo {author} {\bibfnamefont {G.}~\bibnamefont
  {Kresse}}\ and\ \bibinfo {author} {\bibfnamefont {J.}~\bibnamefont
  {Furthm\"uller}},\ }\bibfield  {title} {\enquote {\bibinfo {title} {Efficient
  iterative schemes for ab initio total-energy calculations using a plane-wave
  basis set},}\ }\href {https://doi.org/10.1103/PhysRevB.54.11169} {\bibfield
  {journal} {\bibinfo  {journal} {Phys.~Rev.~B}\ }\textbf {\bibinfo {volume}
  {54}},\ \bibinfo {pages} {11169--11186} (\bibinfo {year} {1996})}\BibitemShut
  {NoStop}%
\bibitem [{\citenamefont {Perdew}, \citenamefont {Burke},\ and\ \citenamefont
  {Ernzerhof}(1996)}]{pbe}%
  \BibitemOpen
  \bibfield  {author} {\bibinfo {author} {\bibfnamefont {J.~P.}\ \bibnamefont
  {Perdew}}, \bibinfo {author} {\bibfnamefont {K.}~\bibnamefont {Burke}},\ and\
  \bibinfo {author} {\bibfnamefont {M.}~\bibnamefont {Ernzerhof}},\ }\bibfield
  {title} {\enquote {\bibinfo {title} {Generalized gradient approximation made
  simple},}\ }\href@noop {} {\bibfield  {journal} {\bibinfo  {journal}
  {Phys.~Rev.~Lett.~}\ }\textbf {\bibinfo {volume} {77}},\ \bibinfo {pages}
  {3865--3868} (\bibinfo {year} {1996})}\BibitemShut {NoStop}%
\bibitem [{\citenamefont {Wang}\ \emph {et~al.}(2021)\citenamefont {Wang},
  \citenamefont {Kingsbury}, \citenamefont {McDermott}, \citenamefont {Horton},
  \citenamefont {Jain}, \citenamefont {Ong}, \citenamefont {Dwaraknath},\ and\
  \citenamefont {Persson}}]{wang+21sr}%
  \BibitemOpen
  \bibfield  {author} {\bibinfo {author} {\bibfnamefont {A.}~\bibnamefont
  {Wang}}, \bibinfo {author} {\bibfnamefont {R.}~\bibnamefont {Kingsbury}},
  \bibinfo {author} {\bibfnamefont {M.}~\bibnamefont {McDermott}}, \bibinfo
  {author} {\bibfnamefont {M.}~\bibnamefont {Horton}}, \bibinfo {author}
  {\bibfnamefont {A.}~\bibnamefont {Jain}}, \bibinfo {author} {\bibfnamefont
  {S.~P.}\ \bibnamefont {Ong}}, \bibinfo {author} {\bibfnamefont
  {S.}~\bibnamefont {Dwaraknath}},\ and\ \bibinfo {author} {\bibfnamefont
  {K.~A.}\ \bibnamefont {Persson}},\ }\bibfield  {title} {\enquote {\bibinfo
  {title} {A framework for quantifying uncertainty in dft energy
  corrections},}\ }\href {https://doi.org/10.1038/s41598-021-94550-5}
  {\bibfield  {journal} {\bibinfo  {journal} {Sci.~Rep.~}\ }\textbf {\bibinfo
  {volume} {11}},\ \bibinfo {pages} {15496} (\bibinfo {year}
  {2021})}\BibitemShut {NoStop}%
\bibitem [{\citenamefont {{Pham Thi}}\ \emph {et~al.}(2015)\citenamefont {{Pham
  Thi}}, \citenamefont {Dumas}, \citenamefont {Bouineau}, \citenamefont
  {Dupin}, \citenamefont {Guéneau}, \citenamefont {Gosse}, \citenamefont
  {Benigni}, \citenamefont {Maugis},\ and\ \citenamefont
  {Rogez}}]{pham+15calphad}%
  \BibitemOpen
  \bibfield  {author} {\bibinfo {author} {\bibfnamefont {T.-N.}\ \bibnamefont
  {{Pham Thi}}}, \bibinfo {author} {\bibfnamefont {J.-C.}\ \bibnamefont
  {Dumas}}, \bibinfo {author} {\bibfnamefont {V.}~\bibnamefont {Bouineau}},
  \bibinfo {author} {\bibfnamefont {N.}~\bibnamefont {Dupin}}, \bibinfo
  {author} {\bibfnamefont {C.}~\bibnamefont {Guéneau}}, \bibinfo {author}
  {\bibfnamefont {S.}~\bibnamefont {Gosse}}, \bibinfo {author} {\bibfnamefont
  {P.}~\bibnamefont {Benigni}}, \bibinfo {author} {\bibfnamefont
  {P.}~\bibnamefont {Maugis}},\ and\ \bibinfo {author} {\bibfnamefont
  {J.}~\bibnamefont {Rogez}},\ }\bibfield  {title} {\enquote {\bibinfo {title}
  {Thermodynamic assessment of the cs-te binary system},}\ }\href
  {https://doi.org/10.1016/j.calphad.2014.10.006} {\bibfield  {journal}
  {\bibinfo  {journal} {Calphad}\ }\textbf {\bibinfo {volume} {48}},\ \bibinfo
  {pages} {1--12} (\bibinfo {year} {2015})}\BibitemShut {NoStop}%
\bibitem [{\citenamefont {Böttcher}\ and\ \citenamefont
  {Kretschmann}(1985)}]{boet-kret85}%
  \BibitemOpen
  \bibfield  {author} {\bibinfo {author} {\bibfnamefont {P.}~\bibnamefont
  {Böttcher}}\ and\ \bibinfo {author} {\bibfnamefont {U.}~\bibnamefont
  {Kretschmann}},\ }\bibfield  {title} {\enquote {\bibinfo {title} {Darstellung
  und kristallstruktur von cste4},}\ }\href
  {https://doi.org/10.1002/zaac.19855230418} {\bibfield  {journal} {\bibinfo
  {journal} {Z.~anorg.~allg.~Chem.}\ }\textbf {\bibinfo {volume} {523}},\
  \bibinfo {pages} {145--152} (\bibinfo {year} {1985})},\ \Eprint
  {https://arxiv.org/abs/https://onlinelibrary.wiley.com/doi/pdf/10.1002/zaac.19855230418}
  {https://onlinelibrary.wiley.com/doi/pdf/10.1002/zaac.19855230418}
  \BibitemShut {NoStop}%
\bibitem [{\citenamefont {Böttcher}\ and\ \citenamefont
  {Kretschmann}(1982)}]{boet+kret82}%
  \BibitemOpen
  \bibfield  {author} {\bibinfo {author} {\bibfnamefont {P.}~\bibnamefont
  {Böttcher}}\ and\ \bibinfo {author} {\bibfnamefont {U.}~\bibnamefont
  {Kretschmann}},\ }\bibfield  {title} {\enquote {\bibinfo {title} {Darstellung
  und kristallstruktur von dicaesiumpentatellurid, cs2te5},}\ }\href
  {https://doi.org/10.1002/zaac.19824910106} {\bibfield  {journal} {\bibinfo
  {journal} {Z.~anorg.~allg.~Chem.}\ }\textbf {\bibinfo {volume} {491}},\
  \bibinfo {pages} {39--46} (\bibinfo {year} {1982})},\ \Eprint
  {https://arxiv.org/abs/https://onlinelibrary.wiley.com/doi/pdf/10.1002/zaac.19824910106}
  {https://onlinelibrary.wiley.com/doi/pdf/10.1002/zaac.19824910106}
  \BibitemShut {NoStop}%
\bibitem [{\citenamefont {Böttcher}(1980)}]{boet80jlcm}%
  \BibitemOpen
  \bibfield  {author} {\bibinfo {author} {\bibfnamefont {P.}~\bibnamefont
  {Böttcher}},\ }\bibfield  {title} {\enquote {\bibinfo {title} {Synthesis and
  crystal structure of rb2te3 and cs2te3},}\ }\href
  {https://doi.org/10.1016/0022-5088(80)90235-0} {\bibfield  {journal}
  {\bibinfo  {journal} {J.~Less~Common~Met.}\ }\textbf {\bibinfo {volume}
  {70}},\ \bibinfo {pages} {263--271} (\bibinfo {year} {1980})}\BibitemShut
  {NoStop}%
\bibitem [{\citenamefont {{de Boer}}\ and\ \citenamefont
  {Cordfunke}(1995)}]{debo-cord95jac}%
  \BibitemOpen
  \bibfield  {author} {\bibinfo {author} {\bibfnamefont {R.}~\bibnamefont {{de
  Boer}}}\ and\ \bibinfo {author} {\bibfnamefont {E.}~\bibnamefont
  {Cordfunke}},\ }\bibfield  {title} {\enquote {\bibinfo {title} {On the
  caesium-rich part of the cs-te phase diagram},}\ }\href
  {https://doi.org/10.1016/0925-8388(95)01666-X} {\bibfield  {journal}
  {\bibinfo  {journal} {J.~Alloys~Compd.~}\ }\textbf {\bibinfo {volume}
  {228}},\ \bibinfo {pages} {75--78} (\bibinfo {year} {1995})}\BibitemShut
  {NoStop}%
\bibitem [{\citenamefont {{de Boer}}\ and\ \citenamefont
  {Cordfunke}(1997)}]{debo-cord97}%
  \BibitemOpen
  \bibfield  {author} {\bibinfo {author} {\bibfnamefont {R.}~\bibnamefont {{de
  Boer}}}\ and\ \bibinfo {author} {\bibfnamefont {E.}~\bibnamefont
  {Cordfunke}},\ }\bibfield  {title} {\enquote {\bibinfo {title} {Thermodynamic
  properties of cs5te3},}\ }\href {https://doi.org/10.1006/jcht.1996.0180}
  {\bibfield  {journal} {\bibinfo  {journal} {J.~Chem.~ Thermodyn}\ }\textbf
  {\bibinfo {volume} {29}},\ \bibinfo {pages} {603--608} (\bibinfo {year}
  {1997})}\BibitemShut {NoStop}%
\bibitem [{\citenamefont {Sangster}\ and\ \citenamefont
  {Pelton}(1993)}]{sang-pelt93jpheq}%
  \BibitemOpen
  \bibfield  {author} {\bibinfo {author} {\bibfnamefont {J.}~\bibnamefont
  {Sangster}}\ and\ \bibinfo {author} {\bibfnamefont {A.~D.}\ \bibnamefont
  {Pelton}},\ }\bibfield  {title} {\enquote {\bibinfo {title} {The cs-te
  (cesium-tellurium) system},}\ }\href {https://doi.org/10.1007/BF02667821}
  {\bibfield  {journal} {\bibinfo  {journal} {J.~Ph.~Equilibria}\ }\textbf
  {\bibinfo {volume} {14}},\ \bibinfo {pages} {246--249} (\bibinfo {year}
  {1993})}\BibitemShut {NoStop}%
\bibitem [{\citenamefont {di~Bona}\ \emph {et~al.}(1996)\citenamefont
  {di~Bona}, \citenamefont {Sabary}, \citenamefont {Valeri}, \citenamefont
  {Michelato}, \citenamefont {Sertore},\ and\ \citenamefont
  {Suberlucq}}]{dibo+96jap}%
  \BibitemOpen
  \bibfield  {author} {\bibinfo {author} {\bibfnamefont {A.}~\bibnamefont
  {di~Bona}}, \bibinfo {author} {\bibfnamefont {F.}~\bibnamefont {Sabary}},
  \bibinfo {author} {\bibfnamefont {S.}~\bibnamefont {Valeri}}, \bibinfo
  {author} {\bibfnamefont {P.}~\bibnamefont {Michelato}}, \bibinfo {author}
  {\bibfnamefont {D.}~\bibnamefont {Sertore}},\ and\ \bibinfo {author}
  {\bibfnamefont {G.}~\bibnamefont {Suberlucq}},\ }\bibfield  {title} {\enquote
  {\bibinfo {title} {Auger and x‐ray photoemission spectroscopy study on
  cs2te photocathodes},}\ }\href {https://doi.org/10.1063/1.363161} {\bibfield
  {journal} {\bibinfo  {journal} {J.~Appl.~Phys.~}\ }\textbf {\bibinfo {volume}
  {80}},\ \bibinfo {pages} {3024--3030} (\bibinfo {year} {1996})},\ \Eprint
  {https://arxiv.org/abs/https://doi.org/10.1063/1.363161}
  {https://doi.org/10.1063/1.363161} \BibitemShut {NoStop}%
\bibitem [{\citenamefont {Yusof}\ \emph {et~al.}(2017)\citenamefont {Yusof},
  \citenamefont {Denchfield}, \citenamefont {Warren}, \citenamefont {Cardenas},
  \citenamefont {Samuelson}, \citenamefont {Spentzouris}, \citenamefont
  {Power},\ and\ \citenamefont {Zasadzinski}}]{yuso+17prab}%
  \BibitemOpen
  \bibfield  {author} {\bibinfo {author} {\bibfnamefont {Z.}~\bibnamefont
  {Yusof}}, \bibinfo {author} {\bibfnamefont {A.}~\bibnamefont {Denchfield}},
  \bibinfo {author} {\bibfnamefont {M.}~\bibnamefont {Warren}}, \bibinfo
  {author} {\bibfnamefont {J.}~\bibnamefont {Cardenas}}, \bibinfo {author}
  {\bibfnamefont {N.}~\bibnamefont {Samuelson}}, \bibinfo {author}
  {\bibfnamefont {L.}~\bibnamefont {Spentzouris}}, \bibinfo {author}
  {\bibfnamefont {J.}~\bibnamefont {Power}},\ and\ \bibinfo {author}
  {\bibfnamefont {J.}~\bibnamefont {Zasadzinski}},\ }\bibfield  {title}
  {\enquote {\bibinfo {title} {Photocathode quantum efficiency of ultrathin
  ${\mathrm{cs}}_{2}\mathrm{Te}$ layers on nb substrates},}\ }\href
  {https://doi.org/10.1103/PhysRevAccelBeams.20.123401} {\bibfield  {journal}
  {\bibinfo  {journal} {Phys.~Rev.~ST~Accel.~Beams}\ }\textbf {\bibinfo
  {volume} {20}},\ \bibinfo {pages} {123401} (\bibinfo {year}
  {2017})}\BibitemShut {NoStop}%
\bibitem [{\citenamefont {Pierce}\ \emph {et~al.}(2021)\citenamefont {Pierce},
  \citenamefont {Bae}, \citenamefont {Galdi}, \citenamefont {Cultrera},
  \citenamefont {Bazarov},\ and\ \citenamefont {Maxson}}]{pier+21apl}%
  \BibitemOpen
  \bibfield  {author} {\bibinfo {author} {\bibfnamefont {C.~M.}\ \bibnamefont
  {Pierce}}, \bibinfo {author} {\bibfnamefont {J.~K.}\ \bibnamefont {Bae}},
  \bibinfo {author} {\bibfnamefont {A.}~\bibnamefont {Galdi}}, \bibinfo
  {author} {\bibfnamefont {L.}~\bibnamefont {Cultrera}}, \bibinfo {author}
  {\bibfnamefont {I.}~\bibnamefont {Bazarov}},\ and\ \bibinfo {author}
  {\bibfnamefont {J.}~\bibnamefont {Maxson}},\ }\bibfield  {title} {\enquote
  {\bibinfo {title} {Beam brightness from cs–te near the photoemission
  threshold},}\ }\href {https://doi.org/10.1063/5.0044917} {\bibfield
  {journal} {\bibinfo  {journal} {Appl.~Phys.~Lett.~}\ }\textbf {\bibinfo
  {volume} {118}},\ \bibinfo {pages} {124101} (\bibinfo {year} {2021})},\
  \Eprint {https://arxiv.org/abs/https://doi.org/10.1063/5.0044917}
  {https://doi.org/10.1063/5.0044917} \BibitemShut {NoStop}%
\bibitem [{\citenamefont {Momma}\ and\ \citenamefont {Izumi}(2011)}]{VESTA}%
  \BibitemOpen
  \bibfield  {author} {\bibinfo {author} {\bibfnamefont {K.}~\bibnamefont
  {Momma}}\ and\ \bibinfo {author} {\bibfnamefont {F.}~\bibnamefont {Izumi}},\
  }\bibfield  {title} {\enquote {\bibinfo {title} {{{\it VESTA3} for
  three-dimensional visualization of crystal, volumetric and morphology
  data}},}\ }\href {https://doi.org/10.1107/S0021889811038970} {\bibfield
  {journal} {\bibinfo  {journal} {Journal of Applied Crystallography}\ }\textbf
  {\bibinfo {volume} {44}},\ \bibinfo {pages} {1272--1276} (\bibinfo {year}
  {2011})}\BibitemShut {NoStop}%
\bibitem [{\citenamefont {Sertore}\ \emph {et~al.}(2000)\citenamefont
  {Sertore}, \citenamefont {Schreiber}, \citenamefont {Floettmann},
  \citenamefont {Stephan}, \citenamefont {Zapfe},\ and\ \citenamefont
  {Michelato}}]{sert+00nimpra}%
  \BibitemOpen
  \bibfield  {author} {\bibinfo {author} {\bibfnamefont {D.}~\bibnamefont
  {Sertore}}, \bibinfo {author} {\bibfnamefont {S.}~\bibnamefont {Schreiber}},
  \bibinfo {author} {\bibfnamefont {K.}~\bibnamefont {Floettmann}}, \bibinfo
  {author} {\bibfnamefont {F.}~\bibnamefont {Stephan}}, \bibinfo {author}
  {\bibfnamefont {K.}~\bibnamefont {Zapfe}},\ and\ \bibinfo {author}
  {\bibfnamefont {P.}~\bibnamefont {Michelato}},\ }\bibfield  {title} {\enquote
  {\bibinfo {title} {First operation of cesium telluride photocathodes in the
  ttf injector rf gun},}\ }\href@noop {} {\bibfield  {journal} {\bibinfo
  {journal} {Nucl.~Instrum.~Methods~Phys.~Res.~A}\ }\textbf {\bibinfo {volume}
  {445}},\ \bibinfo {pages} {422--426} (\bibinfo {year} {2000})}\BibitemShut
  {NoStop}%
\bibitem [{\citenamefont {Prat}\ \emph {et~al.}(2015)\citenamefont {Prat},
  \citenamefont {Bettoni}, \citenamefont {Braun}, \citenamefont {Ganter},\ and\
  \citenamefont {Schietinger}}]{prat+15prab}%
  \BibitemOpen
  \bibfield  {author} {\bibinfo {author} {\bibfnamefont {E.}~\bibnamefont
  {Prat}}, \bibinfo {author} {\bibfnamefont {S.}~\bibnamefont {Bettoni}},
  \bibinfo {author} {\bibfnamefont {H.-H.}\ \bibnamefont {Braun}}, \bibinfo
  {author} {\bibfnamefont {R.}~\bibnamefont {Ganter}},\ and\ \bibinfo {author}
  {\bibfnamefont {T.}~\bibnamefont {Schietinger}},\ }\bibfield  {title}
  {\enquote {\bibinfo {title} {Measurements of copper and cesium telluride
  cathodes in a radio-frequency photoinjector},}\ }\href@noop {} {\bibfield
  {journal} {\bibinfo  {journal} {Phys.~Rev.~ST~Accel.~Beams}\ }\textbf
  {\bibinfo {volume} {18}},\ \bibinfo {pages} {043401} (\bibinfo {year}
  {2015})}\BibitemShut {NoStop}%
\bibitem [{\citenamefont {Parzyck}\ \emph {et~al.}(2021)\citenamefont
  {Parzyck}, \citenamefont {Galdi}, \citenamefont {Nangoi}, \citenamefont
  {DeBenedetti}, \citenamefont {Balajka}, \citenamefont {Faeth}, \citenamefont
  {Paik}, \citenamefont {Hu}, \citenamefont {Arias}, \citenamefont {Hines},
  \citenamefont {Schlom}, \citenamefont {Shen},\ and\ \citenamefont
  {Maxson}}]{parz+21arxiv}%
  \BibitemOpen
  \bibfield  {author} {\bibinfo {author} {\bibfnamefont {C.~T.}\ \bibnamefont
  {Parzyck}}, \bibinfo {author} {\bibfnamefont {A.}~\bibnamefont {Galdi}},
  \bibinfo {author} {\bibfnamefont {J.~K.}\ \bibnamefont {Nangoi}}, \bibinfo
  {author} {\bibfnamefont {W.~J.~I.}\ \bibnamefont {DeBenedetti}}, \bibinfo
  {author} {\bibfnamefont {J.}~\bibnamefont {Balajka}}, \bibinfo {author}
  {\bibfnamefont {B.~D.}\ \bibnamefont {Faeth}}, \bibinfo {author}
  {\bibfnamefont {H.}~\bibnamefont {Paik}}, \bibinfo {author} {\bibfnamefont
  {C.}~\bibnamefont {Hu}}, \bibinfo {author} {\bibfnamefont {T.~A.}\
  \bibnamefont {Arias}}, \bibinfo {author} {\bibfnamefont {M.~A.}\ \bibnamefont
  {Hines}}, \bibinfo {author} {\bibfnamefont {D.~G.}\ \bibnamefont {Schlom}},
  \bibinfo {author} {\bibfnamefont {K.~M.}\ \bibnamefont {Shen}},\ and\
  \bibinfo {author} {\bibfnamefont {J.~M.}\ \bibnamefont {Maxson}},\
  }\href@noop {} {\enquote {\bibinfo {title} {A single-crystal alkali
  antimonide photocathode: high efficiency in the ultra-thin limit},}\ }
  (\bibinfo {year} {2021}),\ \Eprint {https://arxiv.org/abs/2112.14366}
  {arXiv:2112.14366 [physics.acc-ph]} \BibitemShut {NoStop}%
\bibitem [{\citenamefont {Antoniuk}\ \emph {et~al.}(2020)\citenamefont
  {Antoniuk}, \citenamefont {Yue}, \citenamefont {Zhou}, \citenamefont
  {Schindler}, \citenamefont {Schroeder}, \citenamefont {Dunham}, \citenamefont
  {Pianetta}, \citenamefont {Vecchione},\ and\ \citenamefont
  {Reed}}]{anto+20prb}%
  \BibitemOpen
  \bibfield  {author} {\bibinfo {author} {\bibfnamefont {E.~R.}\ \bibnamefont
  {Antoniuk}}, \bibinfo {author} {\bibfnamefont {Y.}~\bibnamefont {Yue}},
  \bibinfo {author} {\bibfnamefont {Y.}~\bibnamefont {Zhou}}, \bibinfo {author}
  {\bibfnamefont {P.}~\bibnamefont {Schindler}}, \bibinfo {author}
  {\bibfnamefont {W.~A.}\ \bibnamefont {Schroeder}}, \bibinfo {author}
  {\bibfnamefont {B.}~\bibnamefont {Dunham}}, \bibinfo {author} {\bibfnamefont
  {P.}~\bibnamefont {Pianetta}}, \bibinfo {author} {\bibfnamefont
  {T.}~\bibnamefont {Vecchione}},\ and\ \bibinfo {author} {\bibfnamefont
  {E.~J.}\ \bibnamefont {Reed}},\ }\bibfield  {title} {\enquote {\bibinfo
  {title} {Generalizable density functional theory based photoemission model
  for the accelerated development of photocathodes and other photoemissive
  devices},}\ }\href {https://doi.org/10.1103/PhysRevB.101.235447} {\bibfield
  {journal} {\bibinfo  {journal} {Phys.~Rev.~B}\ }\textbf {\bibinfo {volume}
  {101}},\ \bibinfo {pages} {235447} (\bibinfo {year} {2020})}\BibitemShut
  {NoStop}%
\bibitem [{\citenamefont {Nangoi}\ \emph {et~al.}(2021)\citenamefont {Nangoi},
  \citenamefont {Karkare}, \citenamefont {Sundararaman}, \citenamefont
  {Padmore},\ and\ \citenamefont {Arias}}]{nang+21prb}%
  \BibitemOpen
  \bibfield  {author} {\bibinfo {author} {\bibfnamefont {J.~K.}\ \bibnamefont
  {Nangoi}}, \bibinfo {author} {\bibfnamefont {S.}~\bibnamefont {Karkare}},
  \bibinfo {author} {\bibfnamefont {R.}~\bibnamefont {Sundararaman}}, \bibinfo
  {author} {\bibfnamefont {H.~A.}\ \bibnamefont {Padmore}},\ and\ \bibinfo
  {author} {\bibfnamefont {T.~A.}\ \bibnamefont {Arias}},\ }\bibfield  {title}
  {\enquote {\bibinfo {title} {Importance of bulk excitations and coherent
  electron-photon-phonon scattering in photoemission from pbte(111): Ab initio
  theory with experimental comparisons},}\ }\href
  {https://doi.org/10.1103/PhysRevB.104.115132} {\bibfield  {journal} {\bibinfo
   {journal} {Phys.~Rev.~B}\ }\textbf {\bibinfo {volume} {104}},\ \bibinfo
  {pages} {115132} (\bibinfo {year} {2021})}\BibitemShut {NoStop}%
\bibitem [{\citenamefont {Schier}(2021)}]{schier-master}%
  \BibitemOpen
  \bibfield  {author} {\bibinfo {author} {\bibfnamefont {R.}~\bibnamefont
  {Schier}},\ }\emph {\bibinfo {title} {An \textit{ab initio} study of
  CsK$_2$Sb surface facets}},\ \href@noop {} {\bibinfo {type} {Master's
  thesis}},\ \bibinfo  {school} {Humboldt-Universit\"at zu Berlin} (\bibinfo
  {year} {2021})\BibitemShut {NoStop}%
\end{thebibliography}

%

\end{document}